\documentclass[aps,prd,preprintnumbers,nofootinbib]{revtex4}
\usepackage{epsfig}

\preprint{SNUTP-03-012,UW/PT 03-09}
\begin{document}
\title{ 
Perturbative matching of staggered four-fermion operators 
with hypercubic fat links}
%
%
%
\author{Weonjong Lee}
\email{wlee@phya.snu.ac.kr}
\affiliation{
  School of Physics,
  Seoul National University,
  Seoul, 151-747, South Korea
  }
\author{Stephen Sharpe}
\email{sharpe@phys.washington.edu}
\affiliation{
  Physics Department,
  University of Washington,
  Seattle, WA 98195-1560, USA 
  }
\date{\today}
\begin{abstract}
We calculate the one-loop matching coefficients between continuum and
lattice four-fermion operators for lattice operators constructed using
staggered fermions and improved by the use of fattened links. In
particular, we consider hypercubic fat links and $SU(3)$ projected
Fat-7 links, and their mean-field improved versions.  We calculate
only current-current diagrams, so that our results apply for operators
whose flavor structure does not allow ``eye-diagrams''.  We present
general formulae, based on two independent approaches, and give
numerical results for the cases in which the operators have the taste
(staggered flavor) of the pseudo-Goldstone pion.  We find that the
one-loop corrections are reduced down to the 10-20\% level, resolving
the problem of large perturbative corrections for staggered fermion
calculations of matrix elements.
\end{abstract}
\pacs{11.15.Ha, 12.38.Gc, 12.38.Bx}
%
%
\maketitle

\section{Introduction}
\label{sec:intr}
It has recently become clear that improving staggered fermions through
the use of fattened links greatly reduces both the breaking of taste
symmetry\footnote{%
We adopt the recently proposed name for what was previously known as
``staggered flavor'' symmetry.  We also refer to ``fattened'' rather
than ``smeared'' links, in order to distinguish our operators from the
smeared operators introduced in Ref.~\cite{ref:sharpe:3}, which
involved smearing of the fermion fields.}
and the size of one-loop matching coefficients.
For example, two of us (WL and SS) calculated the one-loop matching
corrections for all bilinear operators constructed on a hypercube,
and found them to be reduced by fattening from as large as $\sim 50\%$
to the $\sim 10\%$ level for various choices of
fattening~\cite{ref:wlee:3}.  The greatest reduction is for
``Fat-7''~\cite{Toussaint:1998sa} and hypercubic fat (HYP)
links~\cite{ref:anna:0}, the latter after mean-field
improvement.\footnote{%
See Ref.~\cite{DeGrand:2002vu} for an extensive study of the
properties of fattened links for other quantities.}
Based on this, and on the result that HYP links lead to a greater
reduction in taste symmetry breaking in the pion multiplet than other
local fattening choices~\cite{ref:anna:0}, we are undertaking
numerical calculations of weak matrix elements using staggered
fermions with HYP fattened links.
The weak matrix elements include those relevant for predicting CP
violation in the kaon system.
These matrix elements involve four-fermion operators, and an essential
adjunct to the numerical calculations is a determination of the
one-loop matching coefficients for such operators. In this paper we
present the results for these matching coefficients for operators,
constructed using HYP links, in which only ``current-current''
diagrams contribute. This completes one-loop calculation for the
operators relevant for $B_K$ and the $\Delta I=3/2$ component of the
$K\to\pi\pi$ amplitudes.  For the most general four-fermion operator,
one also needs to calculate the contribution of ``penguin'' diagrams,
results of which we hope to present soon.
Calculations with unimproved staggered fermions suggest that 
one-loop corrections for four-fermion operators will be of similar
size to those for the bilinears from which they are constructed.  For
example, for gauge-invariant unimproved four-fermion operators, the
corrections range up to 100\%, roughly twice those for the
corresponding bilinears~\cite{ref:wlee:1}.
Our results confirm this expectation for improved operators.  One-loop
contributions to the matching coefficients are at the 10\% level for
nearly all operators, small enough that the systematic error in
results for matrix elements due to the missing higher loop
contributions will likely be smaller than those from other sources, at
least in the next few years.
This paper is organized as follows. 
In Sec.~\ref{sec:notation+review}, we present the action and composite
operators made of staggered fermions along with fat links, and discuss
the Feynman rules that follow.
In Sec.~\ref{sec:renorm:latt} we discuss the calculation of the
renormalization of the lattice operators, using two independent
methods.  General results are given in four appendices, and the
specific numerical values of renormalization constants for operators
of particular interest are given in Tables
\ref{tab:ff-op-1-I-1}--\ref{tab:ff-op-T-S-P-II-0}.
Matching with continuum operators is discussed in Sec.~\ref{sec:matching},
and we close with some conclusions in Sec.~\ref{sec:conclude}.
Some preliminary results from this paper were presented in
Ref.~\cite{ref:wlee:lat2002}.
This work is done as a part of the staggered $\epsilon'/\epsilon$
project \cite{ref:wlee:lat2002}.

\section{Action, operators and Feynman rules}
\label{sec:notation+review}
%
%
%
%
The calculations required when using fattened links are a
straightforward extension of those needed for unimproved (``naive'')
staggered fermions.  In particular, Feynman rules for the latter, in
the notation we use here, can be found in
Refs.~\cite{ref:smit:0,ref:daniel:0,ref:sharpe:3,ref:wlee:2}, while
matching factors for four-fermion operators with naive staggered
fermions are calculated in
Refs.~\cite{ref:sheard:1,ref:jlqcd:0,ref:sharpe:2,ref:wlee:2,ref:wlee:1}.
Furthermore, many of the additional features introduced by fattening
the links have been presented in the calculation of matching factors
for bilinears~\cite{ref:wlee:3,DeGrand:2002vu}.  In view of this, we
give only a bare-bones summary of the action, operators and Feynman
rules, emphasizing those features special to the present calculation.

The fermion action has the standard staggered form
\begin{eqnarray}
S = a^4 \sum_{n} \bigg[ \frac{1}{2a}
\sum_{\mu} \eta_{\mu}(n)
\Big(
\bar{\chi}(n) V_{\mu}(n) \chi(n + \hat{\mu}) -
\bar{\chi}(n + \hat{\mu}) V^{\dagger}_{\mu}(n) \chi(n)
\Big)
+ m \bar{\chi}(n)\chi(n) \bigg] \ ,
\end{eqnarray}
[where $ n = (n_1,n_2,n_3,n_4)$ is the lattice coordinate and $
\eta_{\mu}(n) = (-1)^{n_1 + \cdots + n_{\mu-1}} $], except that
original ``thin'' links $U_\mu$ are replaced with fattened links
$V_\mu$. 
Note, however, that we continue to use the Wilson plaquette gauge
action constructed out of the thin links.

We use HYP links for the $V_\mu$.  HYP links are defined by three
stages of fattening, and at each stage one must choose a fattening
parameter. We consider two choices of fattening parameters: those
determined by Hasenfratz and Knechtli using a non-perturbative
optimization~\cite{ref:anna:0};
and the values which arise when implementing the Symanzik improvement
program~\cite{ref:lepage:0}. It turns out that the latter choice is
equivalent, in one-loop calculations, to using Fat-7 links if, in
addition, one projects the links back into $SU(3)$ (which we call
$\overline{\rm Fat7}$ links)\cite{ref:wlee:0}.

The key features of HYP links are (i) that they have reduced coupling
to the high-momentum gluons which lead to taste-breaking transitions;
(ii) that they are local in the sense of involving only gauge links
contained in hypercubes connected to the original link; and (iii) that
they lead to (or, in the case of $\overline{\rm Fat7}$ links, are
conjectured to lead to) a large reduction in taste symmetry breaking
in the spectrum.  Several useful properties of HYP and $\overline{\rm
Fat7}$ links have been discussed recently by one of
us~\cite{ref:wlee:0}.

The detailed definitions of $V_\mu$ are given in the original
references and need not be repeated here. What is important for us are
the following properties shared by both choices of fattened links.  The
usual definition of gauge fields
\begin{equation}
U_\mu(x) = \exp 
\big( i a A_\mu ( x \!+\! \frac{\hat\mu}{2}) \big) 
\label{eq:thin:gauge}
\end{equation}
can be extended to the fattened fields, since the latter
live in $SU(3)$:
\begin{equation}
V_\mu(x) = \exp 
\big( i a B_\mu ( x \!+\! \frac{\hat\mu}{2}) \big) \,.
\label{eq:fat:gauge} 
\end{equation}
The ``blocked gauge fields'' $B_\mu$ can be expressed in
terms of the usual gauge fields as
\begin{equation}
B_\mu = \sum_{n=1}^{\infty} B_\mu^{(n)}\,,
\end{equation}
where $B^{(n)}$ contains all terms with $n$ factors of $A$.
In our one-loop calculation we need only $B_\mu^{(1)}$.
Although $B_\mu^{(2)}$ does enter in one-loop ``tadpole'' diagrams,
these contributions vanish, since it follows from
the SU(3) projection that $B_\mu^{(2)}$
has the form of a commutator~\cite{ref:sharpe:3,Bernard:1999kc,ref:wlee:0}.
Thus all we need to know is the coefficient in the linear relation
(written, for later convenience, in momentum space)
\begin{equation}
\widetilde B^{(1)}_\mu (k) =  \sum_\nu h_{\mu\nu}(k) \widetilde A_\nu(k) \,.
\end{equation}
The end result~\cite{ref:wlee:0} is that, to generalize the one-loop
calculations from those for naive staggered fermions (where $V_\mu\to
U_\mu$), we need only replace the gauge-field propagator with
\begin{eqnarray}
\langle \widetilde B^{(1),b}_\mu(k) \widetilde B^{(1),c}_\nu(-k) \rangle &=&
\sum_{\rho,\sigma} h_{\mu\rho}(k) \  h_{\nu\sigma}(-k) \   
	\langle \widetilde A^b_\rho(k) \widetilde A^c_\sigma(-k) \rangle
\nonumber \\
&=& \delta^{bc} \sum_{\rho} h_{\mu\rho}(k) \  h_{\nu\rho}(-k) 
\left[ { \sum_{\beta} \frac{4}{a^2} \sin^{2}( \frac{1}{2} ak_{\beta} ) }
\right]^{-1} \,,
\label{eq:glueprop}
\end{eqnarray}
where $b,c$ are color indices, and the last line holds only
in Feynman gauge, which we use in our calculations.
The simplicity of perturbation theory with HYP or SU(3) projected
links has also been emphasized in Ref.~\cite{DeGrand:2002vu}.

A convenient general form for $h_{\mu\nu}(k)$ is~\cite{ref:wlee:3}
\begin{eqnarray}
h_{\mu\nu}(k) &=& \delta_{\mu\nu} D_\mu(k) + 
(1 - \delta_{\mu\nu}) G_{\mu\nu}(k)
\label{eq:hmunu}\\
D_\mu(k) &=&  1 - d_1 \sum_{\nu\ne\mu} {\bar s}_\nu^2
+ d_2 \sum_{\nu < \rho \atop \nu,\rho\ne\mu}{\bar s}_\nu^2 {\bar s}_\rho^2
- d_3 {\bar s}_\nu^2 {\bar s}_\rho^2 {\bar s}_\sigma^2
- d_4 \sum_{\nu\ne\mu} {\bar s}_\nu^4
\\
G_{\mu\nu}(k) &=&
{\bar s}_\mu {\bar s}_\nu \widetilde G_{\nu,\mu}(k) \\
\widetilde G_{\nu,\mu}(k) &=& d_1
- d_2 \frac{({\bar s}_\rho^2+ {\bar s}_\sigma^2)}{2}
+ d_3 \frac{{\bar s}_\rho^2 {\bar s}_\sigma^2}{3}
+ d_4 {\bar s}_\nu^2 \,.
\end{eqnarray}
Here, the coefficients $d_i$ distinguish different choices
of fat links.\footnote{%
We include $d_4$ for completeness although it is not needed
in our present calculations. It is non-zero for fully $O(a^2)$ 
one-loop improved staggered fermions~\cite{ref:lepage:0}.}
Note that $h_{\mu\nu}(-k)=h_{\mu\nu}(k)$.

We consider here the following choices of coefficients:
\begin{enumerate}
\item[(i)] Unimproved links (naive staggered fermions):
\begin{equation}
d_1 = 0, \quad
d_2 = 0, \quad
d_3 = 0, \quad
d_4 = 0.
\end{equation}
\item[(ii)]
HYP links:
\begin{equation}
d_1 = (2/3)\alpha_1(1+\alpha_2(1+\alpha_3)), \quad
d_2 = (4/3)\alpha_1\alpha_2(1+2\alpha_3), \quad
d_3 = 8 \alpha_1\alpha_2\alpha_3, \quad
d_4 = 0,
\end{equation}
where $\alpha_{1-3}$ are fattening parameters.
One choice was determined
in Ref.~\cite{ref:anna:0} using a non-perturbative optimization
procedure: $\alpha_1=0.75$, $\alpha_2=0.6$ $\alpha_3=0.3$.
This gives
\begin{equation}
d_1 = 0.89\,, \quad
d_2 = 0.96\,, \quad
d_3 = 1.08\,, \quad
d_4 = 0\,,
\end{equation}
and we call these links ``HYP(I)''.
\item[(iii)]
HYP links with one-loop Symanzik-improved coefficients,
i.e. coefficients chosen to remove $O(a^2)$ taste symmetry breaking
couplings at tree level.
This choice gives 
\begin{equation}
d_1 = 1, \quad
d_2 = 1, \quad
d_3 = 1, \quad
d_4 = 0.
\end{equation}
These turn out to be identical to the coefficients
for Fat-7 links, and thus we call these links
``HYP(II)/$\overline{\rm Fat7}$''.
\end{enumerate}

\bigskip

We now turn to the four-fermion operators. We construct these
from standard hypercube bilinears~\cite{ref:klubergstern:1}, 
in which the spin and tastes of the four continuum fermions
are spread over a hypercube.
It is useful to recall the form of the gauge-invariant bilinears:
\begin{eqnarray}
[ S \times F ](y)
& = & \frac{1}{N_f^2} \sum_{A,B}
[\bar{\chi_b}(y+A) \ 
(\overline{ \gamma_S \otimes \xi_F } )_{AB} \ 
\chi_c(y+B)] \  {\cal V}^{bc}(y+A,y+B) \,.
\end{eqnarray}
where $y$ denotes the particular $2^4$ hypercube,
and $A,B$ are ``hypercube vectors'' denoting the
positions within the hypercube.
The normalization factor is chosen so that this operator
goes over, in the continuum limit, to a continuum
bilinear with standard normalization.
The matrices $(\overline{ \gamma_S \otimes \xi_F } )_{AB}$
are standard; see, e.g., Refs.~\cite{ref:daniel:0,ref:wlee:3}.
The spin ($S$) and taste ($F$) of the bilinear  
can each be scalar, $ S $, vector, $ V_\mu $, tensor,
$ T_{\mu\nu} $, axial vector, $ A_\mu $, or pseudoscalar, $ P $.

The only new feature of these operators compared to those used with
unimproved staggered fermions lies in the links used to make them
gauge invariant.  
The factor $ {\cal V}^{bc}(y+A,y+B) $ is constructed by averaging over
all of the shortest paths between $ y+A $ and $ y+B $, and for each
path forming the product of the {\em fattened} gauge links
$V_\mu$. 
%
%
When constructing the operators, we use the same fattened links as in
the action, ensuring the conservation of the current $[V\times S]$.
These are the operators whose one-loop matching factors were found to
be small in Ref.~\cite{ref:wlee:3}.
For four-fermion operators we need to distinguish the two ways of
joining color indices to make gauge invariant operators:%
\footnote{%
We do not consider operators made gauge invariant by
fixing to Landau gauge and leaving out links, since, when one
considers penguin diagrams, these mix with lower dimension operators,
requiring additional non-perturbative
subtractions~\cite{ref:sharpe:2}.  This likely makes them impractical
for general studies of matrix elements.}
\begin{eqnarray}
[ S \times F ] [ S' \times F' ]_{I}(y)
& \equiv  & \frac{1}{N_f^4} \sum_{A,B,C,D}
[\bar{\chi}_b^{(1)}(y+A) \  
(\overline{ \gamma_S \otimes \xi_F } )_{AB} \ 
\chi_c^{(2)}(y+B) ] 
[\bar{\chi}_d^{(3)}(y+C) \ 
(\overline{ \gamma_{S'} \otimes \xi_{F'} } )_{CD} \ 
\chi_e^{(4)}(y+D) ] 
\nonumber \\ & & \hspace*{25mm}
\cdot {\cal V}^{be}(y+A,y+D) \ {\cal V}^{dc}(y+C,y+B)
\label{eq:onetrace}\\ \hspace*{0mm}
[ S \times F ] [ S' \times F' ]_{II}(y)
& \equiv & \frac{1}{N_f^4} \sum_{A,B,C,D}
[\bar{\chi}_b^{(1)}(y+A) \  
(\overline{ \gamma_S \otimes \xi_F } )_{AB} \ 
\chi_c^{(2)}(y+B)] 
[\bar{\chi}_d^{(3)}(y+C) \ 
(\overline{ \gamma_{S'} \otimes \xi_{F'} } )_{CD} \ 
\chi_e^{(4)}(y+D)] 
\nonumber \\ 
& & \hspace*{25mm}
\cdot {\cal V}^{bc}(y+A,y+B) \  {\cal V}^{de}(y+C,y+D)
\label{eq:notrace}
\end{eqnarray}
The subscripts $I$ and $II$ indicate the number of color traces
resulting if each bilinear is contracted with a different external
operator. 
We refer to these two types of operator as one-color-trace and
two-color-trace, respectively.  
The superscripts $(1-4)$ label different flavors ({\em not} tastes) of
staggered fermions---choosing them all different, as we do here,
forbids ``penguin'' diagrams.

Because staggered fermions represent four tastes, there is
considerable redundancy in the transcription of continuum operators
onto the lattice. The choice we have made---the so-called
``two-spin-trace'' operators, to be distinguished from the
one-spin-trace operators of Ref.~\cite{ref:wlee:2}---is that made in
most previous work on staggered weak matrix element
calculations~\cite{ref:sharpe:6,ref:daniel:0}, and in our present
numerical studies~\cite{ref:wlee:lat2002}.

For bilinear operators with HYP links, a further reduction in the size
of one-loop matching contributions was achieved using mean-field (or
``tadpole'') improvement~\cite{ref:wlee:3}.  
Mean-field improvement for HYP links is completely analogous to that
for the usual links~\cite{ref:lepage:1}, which was implemented for
staggered fermions in
Refs.~\cite{ref:sharpe:3,ref:jlqcd:0,ref:wlee:1}.  Links are rescaled
and the fields are renormalized as follows:
\begin{equation}
\chi \rightarrow \psi =\sqrt{u_0} \chi
\qquad
\bar{\chi} \rightarrow \bar{\psi} = \sqrt{u_0}
\bar{\chi}
\qquad
V_{\mu} \rightarrow \tilde{V}_\mu = \frac{ V_{\mu} }{ u_0} \,.
\label{eq:rescale}
\end{equation}
The mean-field scaling factor $u_0$ is determined from
the plaquette composed of fattened links
\begin{equation}
u_0 = \bigg[ \frac{1}{3} {\rm Re} \langle \mbox{Tr} V_{\rm Plaq} \rangle
    \bigg]^{1/4} = 1 - \frac{g^2}{(4\pi)^2} C_F I_{MF} + O(g^4)
\label{eq:u0}
\end{equation}
The integral $I_{MF}$ was called $T^c_{\Delta=2}$ in Ref.~\cite{ref:wlee:3},
and is given by
\begin{equation}
I_{MF} 	= \int_k B \bar{s}_2 
	[ \bar{s}_2 \sum_\alpha h_{1\alpha}(k) h_{1\alpha}(k) 
	- \bar{s}_1 \sum_\alpha h_{1\alpha}(k) h_{2\alpha}(k)  ]
\,.
\label{eq:IMF}
\end{equation}
The numerical values are $ I_{MF} = 0.57826(1)$ for HYP(I) links
and $ I_{MF} = 1.05382(3)$ for HYP(II)/$\overline{\rm Fat7}$ links.
These are substantially smaller than the corresponding factor
for the thin link, $\pi^2$, but are nevertheless significant.

\section{Renormalization of the lattice operators}
\label{sec:renorm:latt}

\begin{figure}
\begin{center}
\epsfysize=6in
\epsfbox{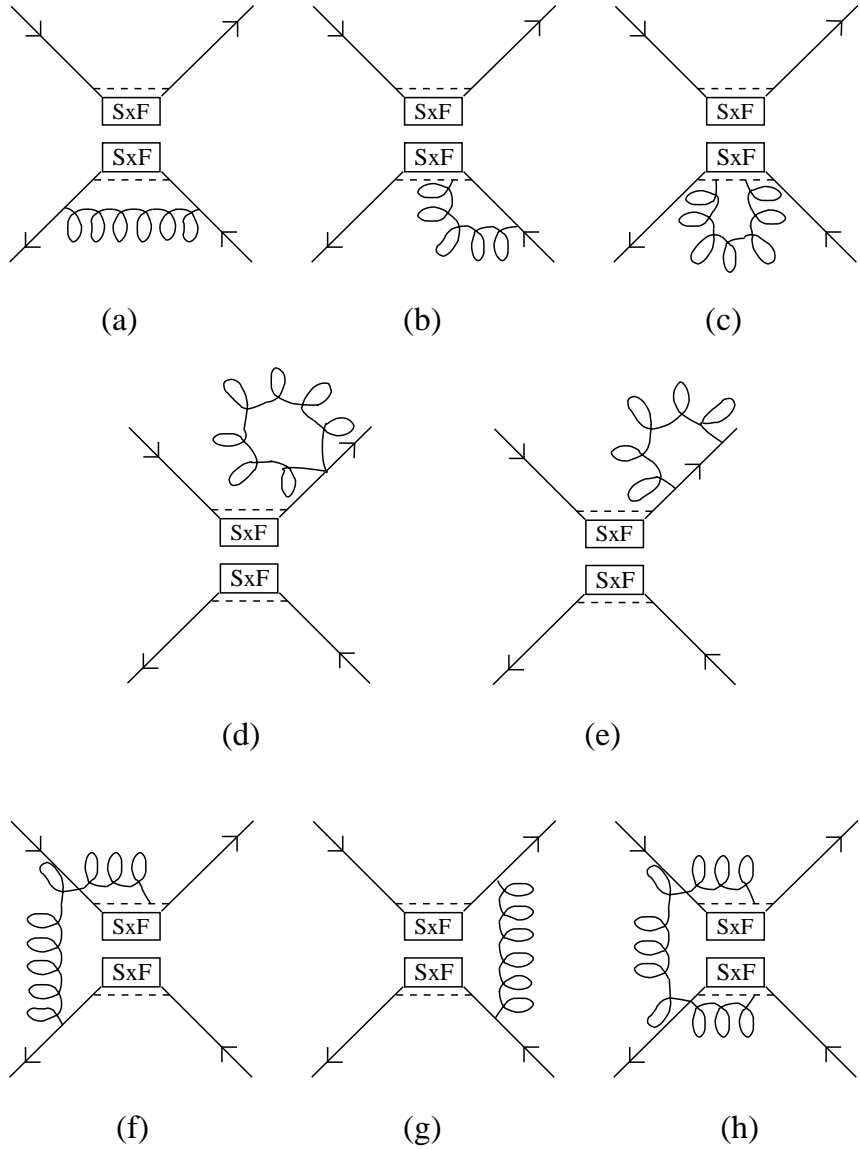}
\end{center}
\caption{Examples of the different types of one-loop diagrams
contributing to matching factors for ``two-color trace'' operators.
The boxes indicate how taste indices are contracted, and the dashed
lines indicate how color indices are contracted.}
\label{fig:twotrace}
\end{figure}

\begin{figure}
\begin{center}
\epsfysize=6in
\epsfbox{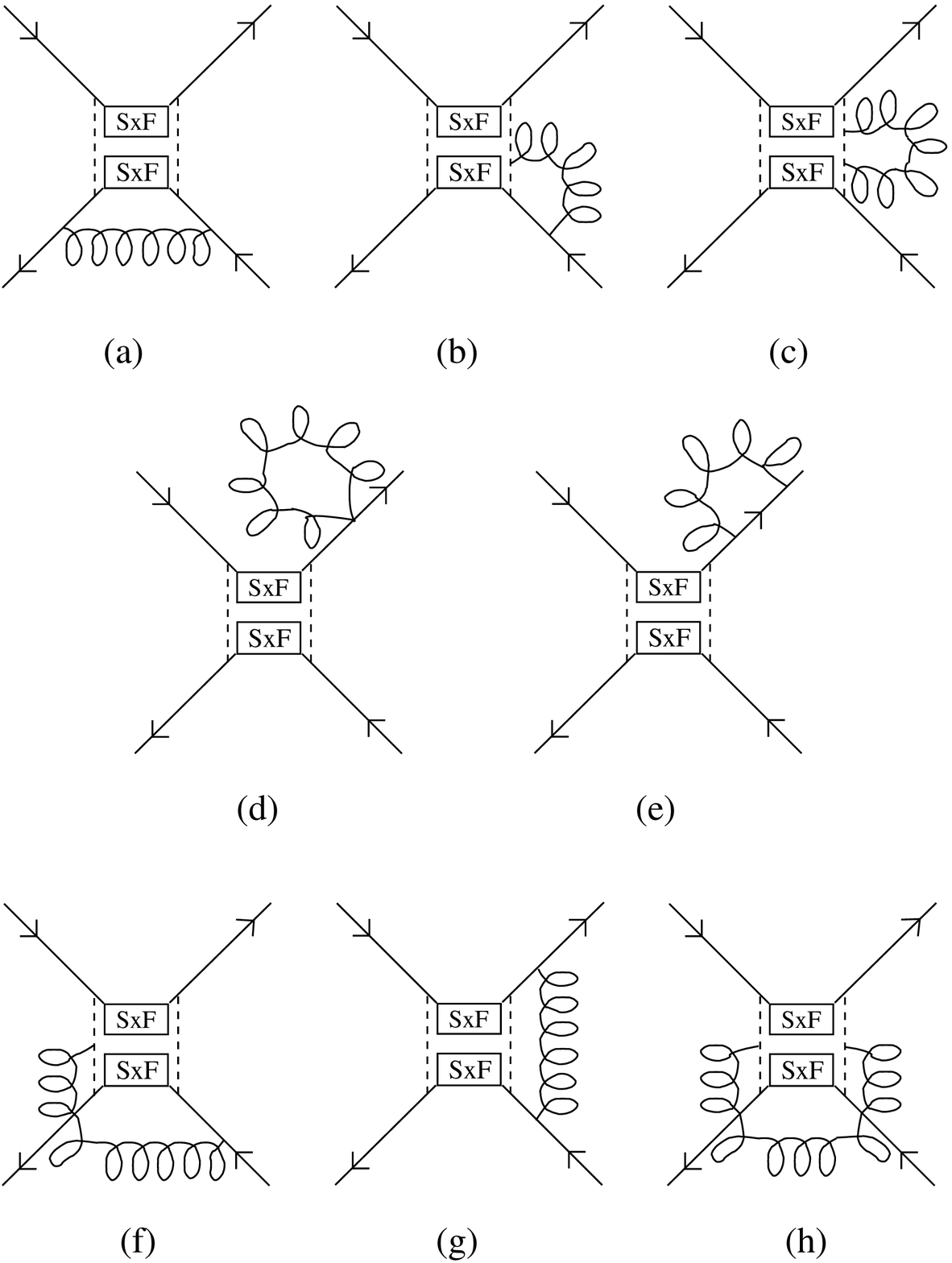}
\end{center}
\caption{Examples of the different types of one-loop diagrams
contributing to matching factors for ``one-color trace'' operators.
The boxes indicate how taste indices are contracted, and the dashed
lines indicate how color indices are contracted.}
\label{fig:onetrace}
\end{figure}

The different types of one-loop diagrams leading to the renormalization
of lattice four-fermion operators with distinct flavors are shown 
in Figs.~\ref{fig:twotrace} and \ref{fig:onetrace}.
These are generically referred to as ``current-current'' diagrams.
Note that the non-locality of the lattice operators leads to many
diagrams in which one or two gluons emanate from the operators
themselves. One of the aims of fattening is to reduce their
contribution, thus making the lattice operators more
``continuum-like''.

The one-loop matrix elements are in general infrared divergent.
Following common practice, we regularize this divergence by
introducing a gluon mass $\lambda$, i.e. we add $\lambda^2$ to the
denominator of the gluon propagator in Eq.~(\ref{eq:glueprop}).  This
allows us to set both the quark masses and the external momenta to
zero.

It is useful to label operators with a 
vector notation~\cite{ref:sharpe:2,ref:wlee:1}
\begin{equation}
\vec{\cal O}^{Latt}_i \equiv \left( \begin{array}{c}
                        {\cal O}^{Latt}_{i,I} \\ {\cal O}^{Latt}_{i,II}
                        \end{array} \right) \,,
\label{eq:vectornotn}
\end{equation}
where $i$ runs over the $16^4$ different choices for
$S$, $S'$, $F$ and $F'$.
This notation reflects the factorization of color factors from
the spin-taste part of the diagram, and is convenient for
expressing the results in a relatively compact way.

The general form of the one-loop matrix element
of the four-fermion operators $\vec{\cal O}^{Latt}_i$
can be written
\begin{equation}
\langle \vec{\cal O}^{Latt}_i\rangle^{(1)} =
\left\{\delta_{ij} + \frac{g^2}{(4\pi)^2} \Big[
        \hat{\gamma}_{ij} \log( a \lambda ) +
        \widehat{C}^{Latt}_{ij} 
	\Big]\right\}
        \langle \vec{\cal O}^{Latt}_j\rangle^{(0)} + O(a)
\label{eq:ff-latt}
\end{equation}
where $ \langle \vec{\cal O}^{Latt}_j\rangle^{(0)}$ is the tree level
matrix element of the $j$'th operator, $\hat\gamma_{ij}$ is the
anomalous dimension matrix\footnote{
The anomalous dimension matrix ($\hat\gamma$) is related to the matrix
($\hat{\Gamma}$) of Ref.~\cite{ref:wlee:1} by $\hat{\Gamma} =
-\hat\gamma$. }
and $\widehat C_{ij}$ is the finite part of the correction.
Both $\hat\gamma$ and $\widehat C$
are matrices in the two-dimensional ``color-index'' space,
as indicated by their ``hats'',
as well as in the explicit indices $i$ and $j$.

Note that, although $\widehat C$ is a very large matrix
[$(16^4)\times(16^4)$ in its $i$, $j$ indices],
we are actually interested only in small blocks within it.
This will be discussed explicitly in the following section.
Nevertheless, we quote the general results since
it takes no more work to do so.

The implementation of mean-field improvement for fattened link
operators follows exactly the same steps as for those with unfattened
links.
The details are given in Ref.~\cite{ref:wlee:1}, and we do not
repeat them here.  The one-loop finite parts are shifted according to
\begin{equation}
\left(\widehat C_{ij}^{Latt}\right)^{MF} 
=\widehat C_{ij}^{Latt} + C_F I_{MF} \widehat T_{ij}
\,,
\label{eq:ff-MF}
\end{equation}
where $\widehat T_{ij}$ are numerical factors which are given below
for the cases of interest, and $I_{MF}$ is given in
Eq.~(\ref{eq:IMF}).

In the Appendices we give the expressions for the contributions to
$\widehat C_{ij}^{Latt}$ for general $i$ and $j$. We have calculated
these using two independent methods.  The first, results of which are
presented in
Appendices~\ref{app:sec:twocolorfact}-\ref{app:sec:onecolor}, is a
generalization of that used in Ref.~\cite{ref:wlee:1,ref:wlee:2}. The
second makes maximal use of previous calculations with bilinears,
reducing the number of additional integrals required.  Results using
this method are presented in Appendix~\ref{app:sec:alt} for mixing in
which all bilinears (before and after mixing) have the same taste $F$,
which are the cases of interest for our numerical calculations.  It is
a non-trivial check of our results that these two methods give results
consistent within the errors of the numerical evaluation of integrals.

We have done two further checks of our results for
$\widehat{C}^{Latt}_{ij}$.
First, we have checked that they are consistent
with the $U(1)_A$ symmetry, which implies that
the renormalization of the
following operators should be identical:
\begin{equation}
[ S \times F ][ S' \times F']\,, \quad
[ S \times F ][ S'5 \times F'5]\,,  \quad
[ S5 \times F5 ][ S' \times F']\,,  \quad
[ S5 \times F5 ][ S'5 \times F'5] \,.
\label{eq:u(1)a}
\end{equation}
Second, we have checked the Fierz identities explained in Appendices
A and B of Ref.~\cite{ref:wlee:1}.
Using these identities, any one-color-trace operator can be represented as a
linear combination of two-color-trace operators and vice versa,
leading to
many non-trivial relations between renormalization constants.

In Tables \ref{tab:ff-op-1-I-1}--\ref{tab:ff-op-T-S-P-II-2}
we give the numerical values for $\hat\gamma_{ij}$, $\widehat C_{ij}^{Latt}$,
and the mean-field factors $\widehat T_{ij}$
for the following five operators:
\begin{eqnarray}
\vec{\cal O}^{Latt}_1 &=&
[ V_\mu \times P ][ V_\mu \times P] + [ A_\mu \times P ][ A_\mu \times P]
\nonumber \\
\vec{\cal O}^{Latt}_2 &=&
[ V_\mu \times P ][ V_\mu \times P] - [ A_\mu \times P ][ A_\mu \times P]
\nonumber \\
\vec{\cal O}^{Latt}_3 &=& -2 \Big(
[ S \times P ][ S \times P] - [ P \times P ][ P \times P] \Big)
\nonumber \\
\vec{\cal O}^{Latt}_4 &=&
[ S \times P ][ S \times P] + [ P \times P ][ P \times P]
\nonumber \\
\vec{\cal O}^{Latt}_5 &=& - \frac{1}{2} \Big(
[ S \times P ][ S \times P] + [ P \times P ][ P \times P]
- \sum_{\mu < \nu} [ T_{\mu\nu} \times P ][ T_{\mu\nu} \times P ]
\Big)\,.
\label{eq:latt-basis}
\end{eqnarray}
Here the color-trace indices are omitted on the right-hand-side.
The first three operators ($\vec{\cal O}^{Latt}_i$, $i=1,2,3$) contribute
to CP-violating transitions of kaons in the standard model; the remaining two
contribute in models of physics beyond the standard model.

The impact of fattening the links can be seen clearly from the table.
Consider first matching between operators with different
tastes (i.e. coefficients in the tables with tastes other than $P$).
Since taste is conserved in the continuum, these coefficients
are the full matching factors. They indicate
the size of taste-breaking due to the lattice action.
As the tables show, these coefficients are typically reduced by
an order of magnitude when the links are fattened.
Taste-conserving coefficients are also reduced substantially, but, since
here there are additional contributions to the matching factors, we postpone
further discussion until the following section.

\section{Matching with continuum operators}
\label{sec:matching}

The continuum operators we wish to match to are:
\begin{eqnarray}
{\cal O}_1 &=&  (\bar{\psi}_1 \gamma_\mu L \psi_2)
                (\bar{\psi}_3 \gamma_\mu L \psi_4)
\nonumber \\
{\cal O}_2 &=&  (\bar{\psi}_1 \gamma_\mu L \psi_2)
                (\bar{\psi}_3 \gamma_\mu R \psi_4)
\nonumber \\
{\cal O}_3 &=& -2 (\bar{\psi}_1 L \psi_2)
                (\bar{\psi}_3 R \psi_4)
\nonumber \\
{\cal O}_4 &=&  (\bar{\psi}_1 L \psi_2)
                (\bar{\psi}_3 L \psi_4)
\nonumber \\
{\cal O}_5 &=&  - \frac{1}{8}
                \Bigl( 1 + \frac{3}{4} \varepsilon \Bigr)
                \sum_{\mu,\nu}
                (\bar{\psi}_1 \gamma_\mu \gamma_\nu L \psi_2)
                (\bar{\psi}_3 \gamma_\nu \gamma_\mu L \psi_4) \,.
\label{eq:cont-basis}
\end{eqnarray}
Here $L,R=1\pm\gamma_5$, and color indices are not shown.
As on the lattice, the operators come in color types I and II.
Matrix elements of the operators are regularized using the 
$\overline{\rm MS}$ scheme, with $\epsilon=(4-d)/2$.
The particular basis we have chosen is the ``practical basis"
of Ref.~\cite{ref:sharpe:4}, in which 
the peculiar factors multiplying ${\cal O}_5$ are required to maintain 
four-dimensional Fierz relations among renormalized operators.
For more discussion on this point, and for the conditions which fully
define the $\overline{\rm MS}$ scheme, see Ref.~\cite{ref:sharpe:4}.

Our aim is to find the lattice operators which, at one-loop level,
match onto the positive parity parts
of these operators.\footnote{We could equally well consider the negative parity parts.
Matching to these is accomplished by considering lattice operators obtained
by multiplying by appropriate factors of $\gamma_5\times \xi_5$.}
To do this we must face the fact that our lattice theory has the additional taste
degree of freedom. We do so in the manner laid out in
Refs.~\cite{ref:sharpe:6,ref:sharpe:1,Kilcup:1997ye,Pekurovsky:1998jd}.
First, we divide each quark loop by a factor of four, corresponding to 
taking the fourth-root of the determinant in simulations with dynamical quarks.
This step is not needed here, since we do not consider penguin contractions
or two-loop diagrams.
Second, we match lattice matrix elements with particular external tastes to
continuum matrix elements. To do this we must divide out by appropriate taste
factors in the lattice matrix elements.\footnote{%
When we consider operators in which two or more of the four flavors are the
same, then, in general,
we have to match each contraction of the continuum operator onto a particular
matrix element of a different lattice operator.}
In this second stage, there is considerable redundancy corresponding to the
freedom to pick different tastes. We have resolved this redundancy by choosing
operators with the taste of the lattice pseudo-Goldstone bosons.

A useful way of thinking about this  matching is the following three stage process.
First, 
we match lattice operators with typical lattice spacings ($a\sim 0.1\,$fm)
onto lattice operators with infinitessimal lattice spacing ($a\to 0$).
This can be done at the level of operators, and is straightforward in principle,
aside from potential theoretical issues arising from taking the fourth-root
of the determinant.
Second, 
we match onto continuum operators defined in our $\overline{\rm MS}$ scheme
at an extremely high scale ($\mu\to\infty$). This matching must be done
between matrix elements (or contractions in general), but it is straightforward
to remove the taste factors since one can work at tree level.
Third,
we match these very high scale continuum operators onto those at a typical
scale, $\mu\sim 2\,$GeV. This is again straightforward.
This three stage process is an adaptation of that laid out by Ji for normal
matching factors~\cite{Ji}. 
We implement it here at the one-loop level.

To carry out the third stage of this process we need the 
one-loop renormalization of the continuum operators defined above.
This has been
worked out in Ref.~\cite{ref:sharpe:4,ref:wlee:1}, and we recall the essential
results here.
Using the same vector notation for the different choices of color
indices defined in Eq.~(\ref{eq:vectornotn}), the renormalized
one-loop matrix elements are
\begin{equation}
\langle\vec{\cal O}_i\rangle^{Cont(1)} =
	\langle\vec{\cal O}_i\rangle^{Cont(0)} + \frac{g^2}{(4\pi)^2} \Big[
        \hat{\gamma}_{ij} \log(\frac{\lambda}{\mu}) +
        \widehat{C}^{Cont}_{ij} 
	\Big]
        \langle\vec{\cal O}_j\rangle^{Cont(0)}
\,.
\label{eq:ff-conti}
\end{equation}
The finite constants can be written as follows:
\begin{equation}
\hat{C}^{Cont}_{ij} = 
\left( \begin{array}{cc}
{-1/6} 	&  {1/2} \\
 {0}   	&  {4/3} 
\end{array} \right) \otimes {\cal M}^a_{ij} +
\left( \begin{array}{cc}
{4/3} 	&  {0} \\
 {1/2}   	&  {-1/6} 
\end{array} \right) \otimes {\cal M}^b_{ij} +
\left( \begin{array}{cc}
{-1/6} 	&  {1/2} \\
 {1/2}   	&  {-1/6} 
\end{array} \right) \otimes {\cal M}^c_{ij} 
\,,
\label{eq:contrenorm}
\end{equation}
where the first matrix in each tensor product acts on the color-trace
indices. The mixing matrices which act on the operator indices are
\begin{eqnarray}
{\cal M}_a &=&
\left( \begin{array}{ccccc}
0	& 0	& 0	& 0 	& 0	\\
0	& 0	& 0	& 0 	& 0	\\
0	& 0	& 5	& 0 	& 0	\\
0	& 0	& 0	& 5 	& -1	\\
0	& 0	& 0	& -2 	& 1	
\end{array} \right)
\\
{\cal M}_b &=&
\left( \begin{array}{ccccc}
0	& 0	& 0	& 0 	& 0	\\
0	& 5	& 0	& 0 	& 0	\\
0	& 0	& 0	& 0 	& 0	\\
0	& 0	& 0	& 1 	& -2	\\
0	& 0	& 0	& -1 	& 5	
\end{array} \right) \,,
\\
{\cal M}_c &=&
\left( \begin{array}{ccccc}
-11	& 0	& 0	& 0 	& 0	\\
0	& -6	& 0	& 0 	& 0	\\
0	& 0	& -6	& 0 	& 0	\\
0	& 0	& 0	& -5 	& -3	\\
0	& 0	& 0	& -3 	& -5	
\end{array} \right)\,.
\end{eqnarray}
We see that the continuum finite constants are typically of magnitude
$\sim5$. 

The matching coefficients are obtained by equating the one-loop matrix
elements calculated using the continuum and lattice regularizations,
after appropriate taste factors have been divided out of the latter.
Although we actually match matrix elements, we present the results in
the form of an operator matching, since this is more familiar.
The result is:
\begin{eqnarray}
\vec{\cal O}^{Cont}_i &=& \sum_{j}
\Biggl[ \delta_{ij}
- \frac{g^2}{(4\pi)^2} \hat{\gamma}_{ij} \log( \mu a )
+ \frac{g^2}{(4\pi)^2} \hat{c}_{ij} 
\Biggr] \vec{\cal O}^{Latt}_j
\label{eq:ff-match} \\
\hat{c}_{ij} &=& \Bigl( \widehat{C}^{Cont}_{ij}
                        - \widehat{C}^{Latt}_{ij} \Bigr)
\end{eqnarray}
The numerical values of matching coefficients can be obtained, for
taste $\xi_5$, by combining the result of Eq.~(\ref{eq:contrenorm})
with those in the Tables.
We close this section by giving some numerical examples of the
matching corrections for the operators of phenomenological importance.
For definiteness, we consider lattice operators with mean-field
improved HYP(II)/$\overline{\rm Fat7}$ links. 
In order to facilitate comparison with previous work, we quote the
same blocks of the matching matrices as given in
Ref.~\cite{Kilcup:1997ye} for specific staggered operators.
%

%
For the first example we quote the square sub-matrix with
indices running over the operators
\begin{equation}
k,l =\left\{ ({\cal O}_1)_{I}, ({\cal O}_1)_{II}, ({\cal O}_2)_{I},
({\cal O}_2)_{II} \right\} 
\,.
\end{equation}
Note that the matching of continuum operators of this form actually 
involves many other lattice operators than these four, but the others
have different tastes, and thus do not contribute to matrix elements
with external lattice pseudo-Goldstone bosons until $O(g^4)$.
We find (with $L=\log( \mu a)$):
\begin{equation}
c_{kl} - \gamma_{kl} L 
=
\left( \begin{array}{cccc}
4.5785 + 2L	& -5.8506 - 6L	& -0.2559 	& 0.7676	\\
-5.8161 -6L	& 4.4209 + 2L	& 0		& 2.4824	\\
-0.2559		& 0.7676	& 11.1493 + 16 L& -3.0345	\\
0		& 2.4824	& -0.1839 + 6 L	& 2.5435 - 2 L	
\end{array} \right)\,,
\end{equation}
where the error in the last digit is approximately $\pm 2$.

The second example is a rectangular submatrix, having indices
\begin{equation}
k= \left\{ \frac12 ({\cal O}_3)_I, \frac12 ({\cal O}_3)_{II} \right\}
\,,
\end{equation}
and
\begin{equation}
l= \left\{ \frac12 ({\cal O}_3)_I, \frac12 ({\cal O}_3)_{II},
({\cal O}_4)_I, ({\cal O}_4)_{II} \right\}
\,.
\end{equation}
Again, these are the only contributions to mixing with
operators of the same taste. 
We find:
\begin{equation}
c_{kl} - \gamma_{kl} L 
=
\left( \begin{array}{cccc}
2.9622 -2L	& -1.0019 + 6 L	 & 0.9403	& -2.8208	\\
-3		& 10.9530 + 16 L & 0		& -7.4011	
\end{array} \right)\,,
\end{equation}
where the error in the last digit is approximately $\pm 2$.

Note that these two matching matrices are sufficient for the
calculation of $B_K$, $B_7^{3/2}$ and $B_8^{3/2}$, for which
the relevant operators are, respectively,
$({\cal O}_1)_I + ({\cal O}_1)_{II}$,
$({\cal O}_2)_{II} + ({\cal O}_3)_{I}$,
and 
$({\cal O}_2)_{I} + ({\cal O}_3)_{II}$.

The numerical values of the matching coefficients should be multiplied
by the factor $\alpha_{\overline{MS}}(1/a)/4\pi$. Taking, as an example, 
$1/a = 2\;$GeV, for which this factor is $\approx 1/66$  we find that a typical
coefficient of $\sim 5$ gives rise to a $\sim 10\%$ correction, while
the largest coefficients are roughly twice this size.  These
corrections are small enough that one-loop perturbation theory is
reasonably convergent, which is not the case for unfattened links.

In making these estimates we are assuming that the appropriate scale
(often called $q^*$) is $\mu \approx 1/a$, so that $L=0$.  This is
reasonable for fattened (and mean-field improved) links which have
reduced couplings to high momentum gluons.  A more detailed test of
this point can, however, be made by calculating $q^*$, an issue which
is subtle for operators with anomalous 
dimensions~\cite{ref:LSinprog,ref:DeGrand:1,ref:Hornbostel:1}.

\section{Conclusion}
\label{sec:conclude}

The main result of this paper is that, by using fattened links, a
long-standing obstacle to studying many weak matrix elements of
interest using staggered fermions has been removed.
In particular, one can use the standard methodology of
Refs.~\cite{ref:sharpe:1,Pekurovsky:1998jd}, which employs
external lattice pseudo-Goldstone bosons, and operators spread
out over the unit hypercube, as long as the links in the operators
and the action are appropriately fattened. 
In this case, the one-loop corrections in matching factors are
typically $\sim 10\%$, and range up to $\sim 20\%$, for all
operators. This is the same size as the corrections for other choices of
fermion discretization, in which one can use ultra-local operators
which do not contain links.  We have shown this for HYP links (or
Fat-7 links including SU(3) projection), but we expect a similar
result to hold for other fattening choices which reduce taste-symmetry
breaking.

Strictly speaking, our calculation is only complete for operators
which do not have penguin contractions. It thus applies for the
phenomenologically interesting quantities $B_K$, $B_7^{3/2}$ and
$B_8^{3/2}$, but only a part for $\Delta I=0$ kaon decays. However,
the remaining contribution from penguin diagrams is small for
unfattened links~\cite{ref:sharpe:2}, and we do not see any reason to
expect this to be changed by fattening. We are presently checking this
expectation.

A particularly encouraging feature of our results is the
across-the-board reduction in mixing of operators with different
tastes.  This provides a thorough test of the idea that, with fattened
links, the staggered fermion operators are much closer to their
continuum counterparts.

\section*{Acknowledgments}
\label{sec:acknowledge}
We would like to thank T.~Bhattacharya, N.~Christ, R.~Gupta, G.~Kilcup
and R.~Mawhinney for their support on the staggered $ \epsilon' /
\epsilon $ project.
This work was supported in part by the BK21 program at Seoul National
University, by the SNU foundation \& Overhead Research fund and by
Korea Research Foundation (KRF) through grant KRF-2002-003-C00033,
and by the US Department of Energy through grant
DE-FG03-96ER40956/A006.

\appendix
\section{Two-color-trace operators: factorizable contributions}
\label{app:sec:twocolorfact}
Here we give the expressions for the contributions to the
renormalization of two-color-trace operators from factorizable
diagrams. Examples of each type of such diagrams
are given in Fig.~\ref{fig:twotrace}(a)-(e).
These contributions are the same as those for bilinears, 
and thus can be extracted from the results given
in Ref.~\cite{ref:wlee:3}. We present them again here, but using
a different notation, that which we use in subsequent appendices 
to express the contributions from non-bilinear-like diagrams.

We use the following abbreviations for lattice integrals:
\begin{eqnarray}
\int_k &\equiv& (4\pi)^2 
\prod_{\mu=1}^{4} \int^{+\pi}_{-\pi} \frac{dk_\mu}{2\pi}
\nonumber \\
s_\mu &\equiv& \sin(k_\mu) \,, 
\qquad
\bar{s}_\mu \equiv \sin(k_\mu / 2)
\nonumber \\
c_\mu &\equiv& \cos(k_\mu) \,,
\qquad
\bar{c}_\mu \equiv \cos(k_\mu / 2)
\nonumber \\
B &\equiv& \frac{1}{ 4 \sum_\mu \bar{s}_\mu^2 + (a \lambda)^2} \,,
\qquad
F \equiv \frac{1}{ \sum_\mu s_\mu^2 }
\nonumber \\
x &\equiv& - 2 \ln(a \lambda) + F_{0000} - \gamma_E + 1
\label{eq:appnot}
\end{eqnarray}
where $F_{0000} = 4.36923(1)$ and $\gamma_E = 0.577216\cdots$.
Here $\lambda$ is the ``gluon mass'' that regulates the infra-red
divergences. We set $a\lambda\to0$ except for the logarithms
contained in $x$.

We consider the renormalization of the two-color-trace
four-fermion operator with general spins and tastes:
$[S\times F][S'\times F']_{II}$. Factorizable diagrams lead only
to two-color-trace operators.
The contribution from corrections to one of the bilinears,
say $[S\times F]$, can be written
\begin{eqnarray}
G_{1(a)} &=& \frac{g^2}{(4\pi)^2} C_F \delta_{ab}\delta_{a'b'}
	\bigg[ 
	\sigma_S \ x \ (\overline{\overline{\gamma_S \otimes \xi_F}})_{CD}
	+ \sum_{M,N} \sum_{\mu,\nu,\beta,\rho}
	X^{\mu\nu, \beta\rho}_{MN}
	( \overline{\overline{
	\gamma_{\mu\beta MSN \rho\nu} \otimes \xi_{MFN} }} )_{CD}
	\bigg]
	(\overline{\overline{\gamma_{S'} \otimes \xi_{F'} }})_{C'D'} 
\\
G_{1(b)}  
	&=& \frac{g^2}{(4\pi)^2} C_F \delta_{ab}\delta_{a'b'}
	\sum_{M,N} \sum_{\mu,\nu,\alpha}
	Y^{\mu\nu, \alpha}_{MN} (S_\nu + F_\nu)_{\rm mod\ 2}
\nonumber \\
& & \hspace*{5mm}
	\bigg[
	( \overline{\overline{
	\gamma_{\nu 5 MSN \alpha\mu} \otimes \xi_{\nu 5 MFN} }} )_{CD}
	+ ( \overline{\overline{
	\gamma_{\mu \alpha \nu 5 MSN } \otimes \xi_{\nu 5 MFN} }} )_{CD}
	\bigg]
	(\overline{\overline{\gamma_{S'} \otimes \xi_{F'}}})_{C'D'} 
\\
G_{1(c)} &=&  \frac{g^2}{(4\pi)^2} C_F \delta_{ab} \delta_{a'b'}\frac{1}{2} 
	\bigg[
	- \Delta_{SF} \cdot T^1_{0000} + T_{\Delta_{SF}}
	\bigg]
	(\overline{\overline{	\gamma_S \otimes \xi_F }})_{CD}	
	(\overline{\overline{\gamma_{S'} \otimes \xi_{F'} }})_{C'D'} 
\\
G_{1(d)} &=& \frac{g^2}{(4\pi)^2} C_F \delta_{ab} \delta_{a'b'}
	\frac{1}{2} T^1_{0000}
	(\overline{\overline{ \gamma_S \otimes \xi_F }})_{CD}
	(\overline{\overline{ \gamma_{S'} \otimes \xi_{F'} }})_{C'D'} 
\\
G_{1(e)} &=& - \frac{g^2}{(4\pi)^2} C_F \delta_{ab}\delta_{a'b'}
	\bigg[
	x + 2 I^{(1)}_1 - 2 I^{(2)}_1 + K^{(2)}_1 
	+ 2 I^{(4)}_1 - K^{(4)}_1 + 6 L^{(4)}_{12}
	\bigg]
	(\overline{\overline{ \gamma_S \otimes \xi_F }})_{CD}
	(\overline{\overline{ \gamma_{S'} \otimes \xi_{F'} }})_{C'D'} 
\,.
\end{eqnarray}
Here we have given the contribution from all diagrams of the
given type. We use the definitions
$\sigma_S=(4,1,0)$ for spins 
$S=(1\ {\rm or\ } \gamma_5,\gamma_\mu\ {\rm or\ } \gamma_\mu\gamma_5,
\sigma_{\mu\nu})$,
$\Delta_{SF} = \sum_{\mu} (S_\mu - F_\mu)_{\rm mod\ 2}$ is
the distance between the quark and anti-quark fields in the 
$[S\times F]$ bilinear and $C_F=4/3$.
The color indices $a,b$ ($a',b'$) and spin-taste indices $C,D$ ($C',D'$)
correspond to the bilinear $[S \times F]$ ($[ S' \times F' ]$).
The matrices $ (\overline{\overline{ \gamma_S \otimes \xi_F }} )_{CD}$
are standard; see, for example, Ref.~\cite{ref:daniel:0}.

There are analogous contributions from diagrams which correct the
$[S'\times F']$ bilinear, and which can be obtained from those above
by the interchanges $abCDSF \leftrightarrow a'b'C'D'S'F'$.

The finite loop integrals appearing in these expressions are
\begin{eqnarray}
X^{\mu\nu, \beta\rho}_{MN} &\equiv&
	\int_k \bigg[ B F^2 
	E_M(k) E_N(-k) \bar{c}_\mu \bar{c}_\nu s_\beta s_\rho 
	\sum_{\lambda} h_{\mu\lambda} h_{\nu\lambda}  
	- \frac{1}{4} \delta_{M,0} \delta_{N,0} 
	\delta_{\mu\nu} \delta_{\beta\rho} B^2 \bigg]
\\
Y^{\mu\nu,\alpha}_{MN} &\equiv&
	\int_k \bigg[ B F ( i \ s_\alpha) \bar{c}_\mu 
	\sum_{\lambda} h_{\mu\lambda} h_{\nu\lambda}
	\frac{1}{12} \sum_{\beta\ne\nu} \sum_{j=1}^4
	E_M ( \theta^{(j)}_{\nu\beta} ) 
	E_N ( -\theta^{(j)}_{\nu\beta} )
	\bigg] 
\\
T^{\mu}_{n} &\equiv&
	\int_k \bigg[ B \exp(i k \cdot n)
	\sum_{\lambda} h_{\mu\lambda} h_{\mu\lambda} \bigg]
	= \int_k \bigg[ B 
	\sum_{\lambda} h_{\mu\lambda}^2
	\prod_\alpha \cos(k_\alpha n_\alpha) \bigg]
\\
T_\Delta &\equiv& 
	\int_k \bigg[ B 2 \bar{s}_\mu \bar{s}_\nu
	\bar{g}_\Delta(k) 
	\sum_{\lambda} h_{\mu\lambda} h_{\nu\lambda} 
	\bigg]
\\
\bar{g}_\Delta(k) &=& 
	( 0,\ 0,\ 1,\ 2 + c_\rho,\ 
	3 + 2 c_\rho + c_\rho c_\sigma ) 
	\quad \mbox{for} \quad \rho\ne\sigma\ne\nu\ne\mu
\end{eqnarray}
where $h_{\mu\nu}(k)$ is given in Eq.~(\ref{eq:hmunu}) of the text, and 
%
%
%
\begin{eqnarray}
E_M(k) &=& \prod_{\mu=1}^{4} \frac{1}{2} 
	\bigg( 
	\exp(-ik_\mu/2 ) + (-1)^{\widetilde{M}_\mu} \exp(+ik_\mu/2) 
	\bigg)
\\
\widetilde{M}_\mu &=& \sum_{\alpha \ne \mu} M_\alpha
\\
\phi & = & \sum_{\rho} \phi_{\rho} \hat{\rho}\ , \\
\theta^{(1)}_{\mu\nu}(\phi) & = &
\frac{1}{2} \phi_{\mu} \hat{\mu},
\\
\theta^{(2)}_{\mu\nu}(\phi) & = &
\frac{1}{2} \phi_{\mu} \hat{\mu}
+ \phi_{\nu} \hat{\nu},
\\
\theta^{(3)}_{\mu\nu}(\phi) & = &
\phi
-\frac{1}{2} \phi_{\mu} \hat{\mu},
\\
\theta^{(4)}_{\mu\nu}(\phi) & = &
\phi
-\frac{1}{2} \phi_{\mu} \hat{\mu}
-\phi_{\nu} \hat{\nu} \ .
\end{eqnarray}
Finally, the finite loop integrals needed for the self-energy diagrams are
\begin{eqnarray}
I^{(1)}_\mu &=& \int_k 
	\bigg[ B F 
	\bar{s}_\mu 
	\sum_\nu \bar{c}_\nu s_\nu
	\sum_{\lambda} h_{\mu\lambda} h_{\nu\lambda} 
	\bigg]
\\
I^{(2)}_\mu &=& \int_k 
	\bigg[ B F 
	c_\mu \bar{c}_\mu^2
	\sum_{\lambda} h_{\mu\lambda}^2 - B^2
	\bigg]
\\
K^{(2)}_\beta &=& \int_k 
	\bigg[ B F
	c_\beta \sum_\mu \bar{c}_\mu^2  
	\sum_{\lambda} h_{\mu\lambda}^2 - 4 B^2
	\bigg]
\\
I^{(4)}_\mu &=& \int_k 
	\bigg[ B F F
	\bar{c}_\mu^2 s_\mu \sin(2k_\mu)
	\sum_{\lambda} h_{\mu\lambda}^2 - \frac{1}{2} B^2
	\bigg]
\\
K^{(4)}_\beta &=& \int_k 
	\bigg[ B F F
	s_\beta \sin(2 k_\beta) 
	\sum_\mu \bar{c}_\mu^2 
	\sum_{\lambda} h_{\mu\lambda}^2 - 2 B^2
	\bigg]
\\
L^{(4)}_{\mu\nu} &=& \int_k 
	\bigg[ B F F
	\bar{c}_\mu \bar{c}_\nu	s_\mu \sin(2 k_\nu)
	\sum_{\lambda} h_{\mu\lambda} h_{\nu\lambda} 
	\bigg]
	\qquad \mbox{where} \quad \mu\ne\nu
\,.
\end{eqnarray}
\section{Two-color-trace operators: Non-factorizable contributions}
\label{app:sec:twocolornonfact}

Here we give expressions for the non-factorizable corrections to 
two-color-trace operators, i.e. those from
Figs.~\ref{fig:twotrace}(f-h).
These diagrams lead to mixing with one-color-trace operators,
and introduce new integrals particular to four-fermion operators.

The results are
%
%
\begin{eqnarray}
G_{1(f)} 
	&=&
	\frac{g^2}{(4\pi)^2} 
	\bigg( - \frac{1}{2 N_c} \delta_{ab} \delta_{a'b'}
	+ \frac{1}{2} \delta_{ab'} \delta_{a'b} \bigg) \cdot
	\sum_{\mu\nu\beta} \sum_{M,N,L} U^{\mu\nu\beta}_{NML} 
	\cdot
\nonumber \\
	& &
	\bigg[
	(S_\nu + F_\nu)_{\rm mod\ 2} 
	( \overline{\overline{
	\gamma_{\nu 5 MSN} \otimes \xi_{\nu5 MFN} }} )_{CD}
	\Big\{
	( \overline{\overline{
	\gamma_{S'L \beta\mu} \otimes \xi_{F'L} }} )_{C'D'} 
	- ( \overline{\overline{
	\gamma_{\mu \beta LS'} \otimes \xi_{L F'} }} )_{C'D'} 
	\Big\}
\nonumber \\
	& &
	+ (S'_\nu + F'_\nu)_{\rm mod\ 2} 
	\Big\{
	( \overline{\overline{
	\gamma_{S L \beta\mu} \otimes \xi_{F L} }} )_{CD} 
	- ( \overline{\overline{
	\gamma_{\mu \beta L S} \otimes \xi_{L F} }} )_{CD} 
	\Big\}
	( \overline{\overline{
	\gamma_{\nu 5 MS'N} \otimes \xi_{\nu 5 MF'N} }} )_{C'D'}
	\bigg]
\\
G_{1(g)} 
	&=&
	- \frac{g^2}{(4 \pi)^2} 
	\bigg( - \frac{1}{2 N_c} \delta_{ab} \delta_{a'b'}
	+ \frac{1}{2} \delta_{ab'} \delta_{a'b} \bigg) \cdot
	\sum_{\mu,\nu,\alpha,\beta} \sum_{M,N} 
	\big(
	X^{\mu\nu,\alpha\beta}_{MN} 
	+ \frac{1}{4} x 
	\delta_{\mu\nu} \delta_{\alpha\beta} 
	\delta_{M,0} \delta_{N,0}
	\big) \cdot 
\nonumber \\
& &
	\bigg[
	( \overline{\overline{
	\gamma_{\nu \alpha M S} \otimes \xi_{M F} }} )_{CD} 
	- ( \overline{\overline{
	\gamma_{S M \alpha \nu} \otimes \xi_{F M} }} )_{CD} 
	\bigg] \cdot \bigg[
	( \overline{\overline{
	\gamma_{\mu \beta N S'} \otimes \xi_{N F'} }} )_{C'D'} 
	- ( \overline{\overline{
	\gamma_{S' N \beta \mu} \otimes \xi_{F' N} }} )_{C'D'} 
	\bigg]
\\
G_{1(h)} &=& 
	- \frac{g^2}{(4 \pi)^2} 
	\bigg( - \frac{1}{2 N_c} \delta_{ab} \delta_{a'b'}
	+ \frac{1}{2} \delta_{ab'} \delta_{a'b} \bigg) \cdot
	\sum_{\mu,\nu} \sum_{M,N,K,L}
	V^{\mu\nu}_{MNKL} \cdot
\nonumber \\
& &
	(S_\mu + F_\mu)_{\rm mod\ 2} \cdot
	(S'_\nu + F'_\nu)_{\rm mod\ 2} \cdot
	( \overline{\overline{
	\gamma_{\mu 5 M S N} \otimes \xi_{\mu 5 M F N} }} )_{CD} 
	( \overline{\overline{
	\gamma_{\nu 5 K S' L} \otimes \xi_{\nu 5 K F' L} }} )_{C'D'} 
\end{eqnarray}
where the finite loop integrals $U^{\mu\nu\beta}_{NML}$ and 
$V^{\mu\nu}_{MNKL}$ are defined as
\begin{eqnarray}
U^{\mu\nu\beta}_{NML} &=& \int_k
	\bigg[
	B F \cdot  i \bar{c}_\mu s_\beta \cdot 
	\sum_{\lambda} h_{\mu\lambda} h_{\nu\lambda} \cdot 
	\frac{1}{12} \sum_{\alpha \ne \nu} \sum_{j=1}^{4}
	E_N ( \theta^{(j)}_{\nu\alpha} ) 
	E_M ( k - \theta^{(j)}_{\nu\alpha} ) 
	E_L ( -k ) 
	\bigg]
\\
V^{\mu\nu}_{MNKL} &=&  \int_k
	\bigg[
	B \cdot \sum_{\lambda} h_{\mu\lambda} h_{\nu\lambda} \cdot
	\frac{1}{12} \sum_{\rho \ne \mu} \sum_{j=1}^{4}
	E_M( k - \theta^{(j)}_{\mu\rho} ) 
	E_N( \theta^{(j)}_{\mu\rho} ) \cdot
	\frac{1}{12} \sum_{\sigma \ne \nu} \sum_{i=1}^{4}
	E_K( - k + \theta^{(i)}_{\nu\sigma} ) 
	E_L( - \theta^{(i)}_{\nu\sigma} )
	\bigg]
\,.
\end{eqnarray}
\section{One-color-trace operators}
\label{app:sec:onecolor}
Finally, we give the results for one-color-trace operators.
The contributions from diagrams of the type of Fig.~\ref{fig:onetrace}(a)
are the same as the corresponding
two-color-trace contributions aside from the color factor:
%
%
\begin{eqnarray}
G_{2(a)} &=& G_{1(a)}\bigg( C_F \delta_{ab} \delta_{a'b'} 
	\rightarrow \big\{ - \frac{1}{2 N_c} \delta_{ab} \delta_{a'b'}
	+ \frac{1}{2} \delta_{ab'} \delta_{a'b} \big\}
	\bigg)\,.
\end{eqnarray}
The self-energy diagrams, Figs.~\ref{fig:onetrace}(d-e) also differ
only by color factors, e.g.
\begin{eqnarray}
G_{2(e)} &=& G_{1(e)}( \delta_{ab} \delta_{a'b'} 
	\rightarrow \delta_{ab'} \delta_{a'b} )
\end{eqnarray}
The remaining diagrams give
\begin{eqnarray}
G_{2(b)} 
	&=& - \frac{g^2}{(4\pi)^2} C_F \delta_{ab'} \delta_{a'b}
	\sum_{\mu,\nu} \sum_{\beta} \sum_{M,N}
	\frac{1}{2} \Big[ Y^{\mu\nu,\beta}_{N [\nu5M]}
	+ Y^{\mu\nu,\beta}_{M [\nu5N]} \Big] \cdot
\nonumber \\ & &
	\Big[
	( \overline{\overline{
	\gamma_{\mu \beta M S} \otimes \xi_{M F} }} )_{CD} 
	( \overline{\overline{
	\gamma_{S' N} \otimes \xi_{F' N} }} )_{C'D'} 
	+ ( \overline{\overline{
	\gamma_{S M \beta \mu} \otimes \xi_{F M} }} )_{CD} 
	( \overline{\overline{
	\gamma_{N S'} \otimes \xi_{N F'} }} )_{C'D'} 
\nonumber \\ & &
	+( \overline{\overline{
	\gamma_{S N} \otimes \xi_{F N} }} )_{CD} 
	( \overline{\overline{
	\gamma_{\mu \beta M S'} \otimes \xi_{M F'} }} )_{C'D'} 
	+ ( \overline{\overline{
	\gamma_{N S} \otimes \xi_{N F} }} )_{CD} 
	( \overline{\overline{
	\gamma_{S' M \beta \mu} \otimes \xi_{F' M} }} )_{C'D'}
	\Big]
\\
G_{2(c)} 
	&=& - \frac{1}{2} \frac{g^2}{(4\pi)^2} 
	\cdot C_F \delta_{ab'} \delta_{a'b} \cdot T^1_{0000}
	\cdot
\nonumber \\
	& & \bigg[
	4 ( \overline{ \overline{
	\gamma_{S} \otimes \xi_{F} }} )_{CD} 
	( \overline{ \overline{
	\gamma_{S'} \otimes \xi_{F'} }} )_{C'D'} 
\nonumber \\
	& & \hspace*{10mm}
	- \frac{1}{2} \sum_{\mu} 
	\big\{ (-1)^{S'_\mu + F'_\mu} + (-1)^{S_\mu + F_\mu} \big\}
	( \overline{\overline{
	\gamma_{\mu 5 S} \otimes \xi_{\mu 5 F} }} )_{CD} 
	( \overline{\overline{
	\gamma_{\mu 5 S'} \otimes \xi_{\mu 5 F'} }} )_{C'D'} 
	\bigg]
\nonumber \\
	& & + \frac{1}{8} \frac{g^2}{(4\pi)^2}
	\cdot C_F \delta_{ab'} \delta_{a'b} \cdot
	\sum_{\mu\ne\nu} \sum_{M} \widetilde{T}^{\mu\nu}_{M} \cdot
\nonumber \\
	& & \bigg[
	\big\{ (-1)^{ \tilde{M} \cdot (S + F) }
	+ (-1)^{\tilde{M} \cdot (S' + F') } \big\}
	( \overline{\overline{
	\gamma_{M S} \otimes \xi_{M F} }} )_{CD} 
	( \overline{\overline{
	\gamma_{M S'} \otimes \xi_{M F'} }} )_{C'D'} 
\nonumber \\
	& & - 2 \big\{ (-1)^{ S_\mu + F_\mu + \tilde{M} \cdot (S + F) } 
	+ (-1)^{ S'_\mu + F'_\mu + \tilde{M} \cdot (S' + F') } \big\}
	( \overline{\overline{
	\gamma_{\mu 5 M S} \otimes \xi_{\mu 5 M F} }} )_{CD} 
	( \overline{\overline{
	\gamma_{\mu 5 M S'} \otimes \xi_{\mu 5 M F'} }} )_{C'D'} 
\nonumber \\
	& & + \big\{ (-1)^{ S_\mu + F_\mu + S_\nu + F_\nu + 
	\tilde{M} \cdot (S + F) } 
	+ (-1)^{ S'_\mu + F'_\mu + S'_\nu + F'_\nu + 
	\tilde{M} \cdot (S' + F') } \big\} \cdot
\nonumber \\
	& & \hspace*{10mm}
	( \overline{\overline{
	\gamma_{\mu \nu M S} \otimes \xi_{\mu \nu M F} }} )_{CD} 
	( \overline{\overline{
	\gamma_{\mu \nu M S'} \otimes \xi_{\mu \nu M F'} }} )_{C'D'} 
	\bigg]
\end{eqnarray}
where the new integral, $\widetilde{T}^{\mu\nu}_M$, is
\begin{eqnarray}
\widetilde{T}^{\mu\nu}_M 	&\equiv&
	\int_k \bigg[
	B \cdot \bar{s}_\mu \bar{s}_\nu
	\cdot \sum_{\lambda} h_{\mu\lambda} h_{\nu\lambda} \cdot
	\Big\{
	\frac{1}{2} \delta_{M,0}
	+ \frac{1}{6} \sum_{j=1}^{3} 
	E_M( - \psi^{(j)}_{\mu\nu} ) E_M( \psi^{(j)}_{\mu\nu} ) 
	\Big\}
	\bigg]
\end{eqnarray}
with, for $\mu \ne \nu \ne \rho \ne \sigma$,
\begin{eqnarray}
\psi^{(1)}_{\mu\nu}(k) &=& k_\rho \hat{\rho}
\nonumber \\
\psi^{(2)}_{\mu\nu}(k) &=& k_\sigma \hat{\sigma}
\nonumber \\
\psi^{(3)}_{\mu\nu}(k) &=& k_\rho \hat{\rho} + k_\sigma \hat{\sigma} \, .
\end{eqnarray}
and
\begin{eqnarray}
G_{2(f)} 
	&=& \frac{g^2}{(4\pi)^2} 
	\bigg( - \frac{1}{2 N_c} \delta_{ab} \delta_{a'b'}
	+ \frac{1}{2} \delta_{ab'} \delta_{a'b} \bigg) \cdot
	\sum_{\mu,\nu,\beta} \sum_{L,M,N} 
	\frac{1}{2} \cdot \bigg\{ 
	U^{\mu\nu,\beta}_{[\nu 5 L]MN} 
	- U^{\mu\nu,\beta}_{[\nu 5 M]LN} 
	\bigg\} \cdot
\nonumber \\
	& & \bigg[
	( \overline{\overline{
	\gamma_{\mu \beta N S L} \otimes \xi_{N F L} }} )_{CD} 
	( \overline{\overline{
	\gamma_{M S'} \otimes \xi_{M F'} }} )_{C'D'} 
	+ ( \overline{\overline{
	\gamma_{S M} \otimes \xi_{F M} }} )_{CD} 
	( \overline{\overline{
	\gamma_{L S' N \beta \mu} \otimes \xi_{L F' N} }} )_{C'D'} 
\nonumber \\
	& & + ( \overline{\overline{
	\gamma_{M S} \otimes \xi_{M F} }} )_{CD} 
	( \overline{\overline{
	\gamma_{\mu \beta N S' L} \otimes \xi_{N F' L} }} )_{C'D'} 
	+ ( \overline{\overline{
	\gamma_{L S N \beta \mu} \otimes \xi_{L F N} }} )_{CD} 
	( \overline{\overline{
	\gamma_{S' M} \otimes \xi_{F' M} }} )_{C'D'} 
	\bigg]
\\
G_{2(g)} 
	&=& \frac{g^2}{(4\pi)^2} 
	\sum_{\mu,\nu} \sum_{\rho,\sigma} \sum_{M,N}
	\bigg\{ X^{\mu\nu,\rho\sigma}_{MN}
	+ \frac{1}{4} x \delta_{\rho\sigma} \delta_{\mu\nu}
	\delta_{M,0} \delta_{N,0} \bigg\} \cdot
\nonumber \\
	& & \bigg[
	C_F \delta_{ab'} \delta_{a'b} \cdot
	\Big\{ 
	( \overline{\overline{
	\gamma_{S M \rho \mu} \otimes \xi_{F M} }} )_{CD} 
	( \overline{\overline{
	\gamma_{\nu \sigma N S'} \otimes \xi_{N F'} }} )_{C'D'} 
\nonumber \\
	& & \hspace*{20mm}
	+ ( \overline{\overline{
	\gamma_{\mu \rho M S} \otimes \xi_{M F} }} )_{CD} 
	( \overline{\overline{
	\gamma_{S' N \sigma \nu} \otimes \xi_{F' N} }} )_{C'D'} 
	\Big\}
\nonumber \\
	& & - \Big( - \frac{1}{2 N_c} \delta_{ab} \delta_{a'b'}
	+ \frac{1}{2} \delta_{ab'} \delta_{a'b}	\Big) \cdot
\nonumber \\
	& & \hspace*{10mm}
	\Big\{
	( \overline{\overline{
	\gamma_{S M \rho \mu} \otimes \xi_{F M} }} )_{CD}
	( \overline{\overline{
	\gamma_{S' N \sigma \nu} \otimes \xi_{F' N} }} )_{C'D'}
	+ ( \overline{\overline{
	\gamma_{\mu \rho M S} \otimes \xi_{M F} }} )_{CD}
	( \overline{\overline{
	\gamma_{\nu \sigma N S'} \otimes \xi_{N F'} }} )_{C'D'}	 
	\Big\}
	\bigg]
\\
G_{2(h)} &=& - \frac{g^2}{(4\pi)^2}
	\bigg( - \frac{1}{2 N_c} \delta_{ab} \delta_{a'b'}
	+ \frac{1}{2} \delta_{ab'} \delta_{a'b}	\bigg) \cdot
	\sum_{\mu,\nu} \sum_{M,N} \sum_{K,L} 
	\frac{1}{4} 
	( \overline{\overline{
	\gamma_{L S M} \otimes \xi_{L F M} }} )_{CD} 
	( \overline{\overline{
	\gamma_{N S' K} \otimes \xi_{N F' K} }} )_{C'D'} \cdot
\nonumber \\
	& & \hspace*{20mm}
	\bigg[
	V^{\mu\nu}_{K [\mu 5 L] M [\nu 5 N]}
	+ V^{\mu\nu}_{L [\mu 5 K] N [\nu 5 M]}
	- V^{\mu\nu}_{L [\mu 5 K] M [\nu 5 N]}
	- V^{\mu\nu}_{K [\mu 5 L] N [\nu 5 M]}
	\bigg]
\,.
\end{eqnarray}
\section{An alternative method for calculating renormalization factors}
\label{app:sec:alt}

We have done a second, independent, calculation in order to check our results.
The second method originates from an idea of 
Martinelli~\cite{ref:martinelli:0},
and has been generalized to Landau-gauge staggered four-fermion
operators in Ref.~\cite{ref:sharpe:2}, and general local operators
in Ref.~\cite{ref:sharpe:4}.
Here we generalize it further to staggered gauge invariant operators.
The idea is to use the information available from 
the renormalization of bilinear operators as fully as possible.
It turns out that only two types of diagram are specific to 
four-fermion operators, and these one cannot avoid calculating.

Although this method allows a full calculation of mixing, we have
only done the calculation for specific cases.
We consider initial operators in which
the tastes of the two component bilinears are the same ($F'_i=F_i$),
and calculate their mixing only into operators having the
same tastes $F'_f=F_f=F_i$.
We call this ``taste-diagonal'' mixing.
This is the subset of mixing coefficients that
are needed in our numerical calculations,
in which we use external states of definite taste.
It is legitimate to exclude mixing with
``taste off-diagonal'' operators because this
leads to an error of $O(g^4)$, the same size
as the error we are making anyway by using 1-loop matching factors.

The restriction to taste-diagonal mixing 
allows a number of integrals to be dropped,
and allows us to obtain relatively compact
expressions for the final result. 

The basic observation is that almost all diagrams contributing
to renormalization of lattice four-fermion operators
have already been calculated {\em as part of}
the matching factors for bilinears.
The qualifier ``{\em as part of}'' is important: one needs to know the
results for individual subdiagrams.
To explain this in more detail we use the notation of 
Ref.~\cite{ref:sharpe:3,ref:sharpe:2} for 
the different classes of diagrams: 
X diagrams involve gluon exchange between external quark lines
[e.g. Figs.~\ref{fig:twotrace}(a,g) and \ref{fig:onetrace}(a,g)]; 
Y diagrams involve gluon exchange between external quark lines
and the links in the operator
[e.g. Figs.~\ref{fig:twotrace}(b,f) and \ref{fig:onetrace}(b,f)]; 
Z diagrams are self-energy corrections excluding tadpoles
[e.g. Figs.~\ref{fig:twotrace}(e) and \ref{fig:onetrace}(e)]; 
ZT diagrams are tadpole self-energy corrections
[e.g. Figs.~\ref{fig:twotrace}(d) and \ref{fig:onetrace}(d)]; 
and T diagrams are tadpoles in which the gluon begins and
ends on (in general different) links within in the operator
[e.g. Figs.~\ref{fig:twotrace}(c,h) and \ref{fig:onetrace}(c,h)]. 
This classification applies to corrections for both bilinear
and four-fermion operators.

The self-energy diagrams, both types Z and ZT, 
are independent of the operator under consideration, 
and thus can be taken over from
the bilinear calculation without change.

Contributions from X diagrams 
can also be expressed in terms of bilinear X diagram corrections,
using Fierz transformations and charge conjugation.
This has been explained in Ref.~\cite{ref:sharpe:2}, and we
do not repeat this discussion here. 
There is it also shown how the X, Z and ZT diagrams come
with color factors such that they combine as for the
bilinear calculations.

The Y and T diagrams divide into three classes.
The first, exemplified by Figs.~\ref{fig:twotrace}(b,c), 
contribute to the renormalization
of the individual bilinears in the color two-trace form.
The second, exemplified by Figs.~\ref{fig:onetrace}(b,c), 
contribute to the renormalization of bilinears
after Fierzing the one color trace four fermion operator
into two color trace form.
These two sets of diagrams can be combined with 
the X, Z and ZT diagrams in the way described in the 
following subsection. They do not require calculations beyond
those neede for bilinears.

The third set of Y and T diagrams are intrinsic to four-fermion
operators. For two-color-trace operators
these are exemplified by Figs.~\ref{fig:twotrace}(f,h); these, however,
can be shown to contribute only to taste off-diagonal mixing.
Thus for two-color-trace operators
the bilinear calculations are sufficient.
For one-color-trace operators 
the diagrams exemplified by Figs.~\ref{fig:onetrace}(f,h); 
these, however, do contribute to taste-diagonal mixing and 
have to be calculated explicitly. 
We discuss these calculations in subsections~\ref{app:subsec:crossedtad}
[Fig.~\ref{fig:onetrace}(h)] and
\ref{app:subsec:crossedY} [Fig.~\ref{fig:onetrace}(f)], respectively.
Note that these two contributions are not affected by tadpole
improvement.

\subsection{Bilinear-like diagrams}
\label{app:subsec:bilinearlike}

Here we discuss how the ``bilinear-like'' diagrams can be
calculated using input only from calculations of matching
factors for bilinears.

First note that we only consider initial four-fermion
operators, $\vec{O}_i^{Latt}$,
in which both bilinears have the same spin and taste.
Our restriction to mixing which is taste-diagonal turns out
to imply mixing only with operators of the same type,
i.e. the spin may change, but it does so for both bilinears
in the same way.
The label $i$ of such operators obviously also labels
the spin-taste of the individual bilinears.
The taste-diagonal part of the bilinear corrections,
which is all that we need, are
denoted by $X_i$, $Y_i$, $Z$, $ZT$ and $T_i$ for the
various diagrams. The convention here is that the bilinear color factor
($C_F=4/3$) and a factor of $g^2/(4\pi)^2$ are not included.
The self-energy contributions, $Z$ and $ZT$,
do not have a label since they are the same for all operators.
We also need the Fierz matrix for such four-fermion operators,
${\cal F}_{ij}$, and the charge conjugation matrix, ${\cal C}_{ij}$.
The calculation of both is straightforward and is discussed in 
Ref.~\cite{ref:sharpe:2}.

To express our results we first define two useful matrices
\begin{eqnarray}
{\cal B}_{ij} &=& \delta_{ij} (X_i + Y_i + Z + ZT + T_i) \,;\\
{\cal B}'_{ij} &=& \delta_{ij} (X_i + Z + ZT) \,.
\end{eqnarray}
The former has the full bilinear corrections along the
diagonal; in the latter the
contributions involving the gauge links in the operator are removed.
The result is then
\begin{eqnarray}
\widehat C_{ij}^{Latt}(\mathrm{ bilin. part}) &=&
2 \times
\left( \begin{array}{cc}
(-1/6) 	&  (1/2) \\
0   	&  (4/3) 
\end{array} \right)
\left( \begin{array}{cc}
{\cal B}'_{ij} 	&  0\\
0   	&  {\cal B}_{ij}   
\end{array} \right)
\nonumber\\
&&\mbox{}
+ 2 \times
\left( \begin{array}{cc}
(4/3) & 0\\
(1/2) & (-1/6)
\end{array} \right)
\left( \begin{array}{cc}
{\cal F}_{ik}{\cal B}_{kl}{\cal F}_{lj}  	& 0 \\
0 &  {\cal F}_{ik}{\cal B}'_{kl}{\cal F}_{lj} 	
\end{array} \right)
\nonumber \\
&&\mbox{} 
-2 \times
\left( \begin{array}{cc}
(-1/6) & (1/2) \\
(1/2) & (-1/6)
\end{array} \right)
\times
{\cal C}_{ik}{\cal F}_{kl}{\cal B}'_{lm}{\cal F}_{mn}{\cal C}_{nj}  
\,.
\end{eqnarray}
The factors of two are from the presence of two diagrams
of each type. The color matrices are taken from 
Ref.~\cite{ref:sharpe:2}. 

Expressions for the bilinear contributions can be determined
from Ref.~\cite{ref:wlee:3} as follows.
The $X_i$ are related to the mixing matrix $X_{jk}$ of 
Ref.~\cite{ref:wlee:3} by
\begin{equation}
X_i = X_{ii} + x \ \sigma_S
\,.
\end{equation}
where $x$ is defined in Eq.~\ref{eq:appnot}, and
$\sigma_S=(4,1,0)$ for spins 
$S=(1\ {\rm or}\ \gamma_5,
\gamma_\mu\ {\rm or}\ \gamma_\mu\gamma_5,
\sigma_{\mu\nu})$.
The $Y_i$ are exactly as given in Ref.~\cite{ref:wlee:3}: 
the result depends only on
the distance $\Delta_i$ of the bilinear and is labelled $Y_{\Delta_i}$.
The $T_i$ and $ZT$ are combined in Ref.~\cite{ref:wlee:3}, and given in terms
of two quantities $T^{a}_{\Delta_i}$ and $T^{b}_{\Delta_i}$, 
which also depend only on the distance.
The combinations we want are
\begin{eqnarray}
T_i &=& T^{a}_{\Delta_i+1} + T^{b}_{\Delta_i}
= \Delta_i T^{a}_{\Delta=2} + T^{b}_{\Delta_i}
\,,\\
ZT &=& T^{a}_{\Delta=0}
\,.
\end{eqnarray}
Finally, the non-tadpole self-energy is given by
\begin{equation}
Z = -X_{(\gamma_\mu\otimes 1)} - Y_{\Delta=1}
\,.
\end{equation}

Mean-field improvement is simple to implement in this method: one
simply applies it to the bilinear corrections and then uses the same
formulae. This has the following effects:
\begin{eqnarray}
T_i &\to & T_i + T^c_{\Delta_i+1}
= T_i + \Delta_i T^c_{\Delta=2}
= T_i + \Delta_i I_{MF}
\,,\\
ZT &\to & ZT + T^c_{\Delta=0}
= ZT - I_{MF}
\,,
\end{eqnarray}
where $T^c_{\Delta}$ is given in Ref.~\cite{ref:wlee:3}.

\subsection{Crossed tadpole diagrams}
\label{app:subsec:crossedtad}

Here we report the contribution of Fig.~\ref{fig:onetrace}(h) 
to taste-diagonal mixing.
In fact, one can see that the taste-diagonal contributions are also 
diagonal in the spins of the bilinears, and thus are completely diagonal. 
Following Ref.~\cite{ref:wlee:3}, we first define
\begin{eqnarray}
P_\mu(k) &\equiv& D_\mu(k)^2 + \sum_{\nu\mu} G_{\nu,\mu}(k)^2
\,, \\
4 \bar s_\mu \bar s_\rho O_{\mu\rho} &\equiv&
D_\mu G_{\mu,\rho} + D_\rho G_{\rho,\mu}
+ \sum_{\nu\ne(\mu,\rho)} G_{\nu,\mu}G_{\nu,\rho}
\,,
\end{eqnarray}
which are, respectively, the diagonal and off-diagonal
parts of the propagator from unfattened link to unfattened link.
The integrals that arise are then
\begin{eqnarray}
T_{Cr}^P(|\delta|) &=& \frac{1}{16} \sum_H \sum_\mu (-)^{\delta_\mu} H_\mu
\int_k B(k) P_\mu(k)\; V_\mu(H,\delta,k)\; V_T(H,\delta,k)
\,, \label{eq:crossedTP} \\
T_{Cr}^O(|\delta|) &=& \frac{1}{16} \sum_H \sum'_{\mu,\nu,\rho} 
(-)^{\delta_\mu+\delta_\nu+\delta_\rho} H_\mu H_\nu H_\rho
\frac19 \int_k B(k) 4\; O_{\mu\nu} 
(\bar s'_\mu \bar s'_\nu \bar s'_\rho \bar c'_\sigma)^2\;
V_T(H,\delta,k)
\,, \label{eq:crossedTO}
\end{eqnarray}
where the notation is as follows:
\begin{itemize}
\item
The hypercube vector $\delta$ is $\delta=_2 S-F$,
where $S$ and $F$ are the hypercube vectors representing
the spin and taste of each bilinear in the four-fermion operator.
\item
The indices $\mu$, $\nu$, $\rho$ and $\sigma$ are unequal---a
point reinforced by the prime on the sum in $T_C^O$.
\item
$H$ is a hypercube vector, representing
the displacement (mod-2) between the quark in one bilinear
and the antiquark in the other. For one-color trace
operators, the gauge links span this distance.
When one considers the four-fermion operators, $H$ is averaged over,
as the expression indicates.
\item
To write the result in a way which applies to all $H$,
we have used the notation
\begin{equation}
\bar c'_\mu = \cos(H_\mu k_\mu/2) \,,\qquad
\bar s'_\mu = \sin(H_\mu k_\mu/2) = H_\mu \bar s_\mu \,.
\end{equation}
Note in particular that $s'_\mu=0$ if $H_\mu=0$.
\item
The ``vertex factor'' $V_\mu$ arises from the product
of links of total displacement $H$ from which
the gluon emanates---one on each end of the gluon propagator:
\begin{equation}
V_\mu(H,\delta,k) = (\bar c'_\nu \bar c'_\rho \bar c'_\sigma)^2 
+ \frac19
\left[ (\bar c'_\nu \bar s'_\rho \bar s'_\sigma)^2 
(-)^{\delta_\rho + \delta_\sigma} + {\rm perms}\right] \,.
\label{eq:Vmu}
\end{equation}
\item
Finally, the ``transverse'' form factor $T$ contains dependence
on momenta perpendicular to $H$:
\begin{equation}
V_T(H,\delta,k) = \prod_{\epsilon=1,4} 
\cos([1-H_\epsilon]\delta_\epsilon k_\epsilon)
\,.
\end{equation}
\item
Due to the permutation symmetry between indices,
these integrals infact depend only
on the ``distance'' $|\delta|$ of each bilinear,
so that there are five independent integrals.
\end{itemize}

Adding in the color factor resulting from one gluon exchange,
we find that the contribution to the mixing coefficients
from these diagrams is
\begin{equation}
\widehat C_{ii}^{Latt}(\mathrm{Crossed\ tadpoles}) =
\left( \begin{array}{cc}
-1/6 & 1/2 \\
0   &  0   
\end{array} \right)
\left[T_{Cr}^P(|\delta_i|) + T_{Cr}^O(|\delta_i|)\right]
\,.
\end{equation}

This contribution is not affected by mean-field improvement.

\subsection{Crossed Y diagrams}
\label{app:subsec:crossedY}

The calculation for the crossed Y diagrams, 
exemplified by Fig.~\ref{fig:onetrace}(f), is more involved
than that for the crossed tadpoles. In the end, however,
if one considers only taste-diagonal contributions,
the final form is similar to that for the crossed tadpoles.
It can be written in terms of the two integrals
\begin{eqnarray}
Y_{Cr}^P(|\delta|) &=& \frac{1}{16} \sum_H \sum_\mu (-)^{\delta_\mu} H_\mu
\int_k B(k) F(k) P_\mu(k)\; s_\mu^2\; V^Y_\mu(H,\delta,k)\; V_T(H,\delta,k)
\,, \label{eq:crossedYP} \\
Y_{Cr}^O(|\delta|) &=& \frac{1}{16} \sum_H \sum'_{\mu\nu}
(-)^{\delta_\mu} H_\mu
\int_k B(k) F(k) 4\; O_{\mu\nu} (\bar s'_\mu)^2 s_\nu^2\;
V^Y_\mu(H,\delta,k)\; V_T(H,\delta,k)
\,. \label{eq:crossedYO}
\end{eqnarray}
where the notation is as for the crossed tadpoles,
except that the vertex factor changes to
\begin{equation}
V^Y_\mu(H,\delta,k) = (\bar c'_\nu \bar c'_\rho \bar c'_\sigma)^2 
+ \frac13
\left[ (\bar c'_\nu \bar s'_\rho \bar s'_\sigma)^2 
(-)^{\delta_\rho + \delta_\sigma} + {\rm perms}\right] \,,
\end{equation}
in which the second factor is three times larger than in $V_\mu$,
Eq.~\ref{eq:Vmu}.
Note that the sum over $\nu$ in the expression for $Y_{Cr}^O$
is constrained to avoid $\mu$, but is otherwise free.
In particular, the absence of a prime on $s_\nu^2$ means
that there is a contribution even if $H_\nu$ vanishes.

Adding in the color factor resulting from one gluon exchange,
the overall factor of 2 because the gluon can originate from
either link factor, and including the overall sign,
we find that the contribution to the mixing coefficients
from these diagrams is
\begin{equation}
\widehat C_{ii}^{Latt}(\mathrm{Crossed\ Y's}) =
\left( \begin{array}{cc}
-1/6 & 1/2 \\
0   &  0   
\end{array} \right)
\times (-2) \times \left[Y_{Cr}^P(|\delta_i|) + Y_{Cr}^O(|\delta_i|)\right]
\,.
\end{equation}
Again the result depends only of the distance $|\delta|$.

This contribution is not affected by mean-field improvement.

%
%



%
%

\clearpage

\begin{table}[ht]
\caption{Renormalization constants for
${\cal O}_i =({\cal O}^{Latt}_1)_{I}=
[ V_\mu \times P ][ V_\mu\times P]_{I} + 
[ A_\mu \times P ][ A_\mu \times P]_{I}$.
The anomalous dimension matrix $\hat\gamma_{ij}$
and the finite constants $\hat C^{Latt}_{ij}$ are
defined in Eq.~(\ref{eq:ff-latt}). The coefficients
$\hat T_{ij}$ are needed for mean-field improvement,
as defined in Eq.~(\ref{eq:ff-MF}). All Greek indices are
implicitly summed, with the condition that they are unequal.
Results are accurate to $\pm 2$ in the last digit quoted.
}
\label{tab:ff-op-1-I-1}
\begin{center}
\begin{tabular}{ l c c c c c c }
${\cal O}^{Latt}_j $ & color trace	& $\hat{\gamma}_{ij}$
	& $\hat{C}^{Latt}_{ij}$ & $\hat{C}^{Latt}_{ij}$ 
	& $\hat{C}^{Latt}_{ij}$ & $ \hat T_{ij} $ \\ 
	&	&
	& NAIVE 	& HYP(I)	& $HYP(II)/\overline{\rm Fat7}$
	& 	\\
\hline								     
$[ S \times V_\mu ][ S \times V_\mu ] $ & I  &  0  & $-14.395$ & $-1.756$ & $-2.391$ & $+1$  \\
$[ S \times V_\mu ][ S \times V_\mu ] $ & II &  0  & $-3.992$  & $-0.712$ & $-0.615$ & 0 \\
$[ S \times V_\mu ][ S \times V_\nu ] $ & I  &  0  & $-1.110$  & $-0.238$ & $-0.194$ & 0 \\
$[ S \times A_\mu ][ S \times A_\mu ] $ & I  &  0  & $-0.313$  & $+0.011$ & $+0.071$ & 0 \\
$[ S \times A_\mu ][ S \times A_\mu ] $ & II &  0  & $-1.219$  & $-0.034$ & $-0.061$ & 0 \\
$[ S \times A_\mu ][ S \times A_\nu ] $	& I  &  0  & $-0.119$  & $-0.003$ & $-0.008$ & 0 \\
\hline
\end{tabular}
\end{center}
\end{table}
\begin{table}[ht]
\caption{Renormalization constants for 
$({\cal O}^{Latt}_1)_{I}$
(continued from Table~\ref{tab:ff-op-1-I-1}).}

\label{tab:ff-op-1-I-2}
\begin{center}
\begin{tabular}{ l c c c c c c }
${\cal O}^{Latt}_j $ & color trace	& $\hat{\gamma}_{ij}$
	& $\hat{C}^{Latt}_{ij}$ & $\hat{C}^{Latt}_{ij}$ 
	& $\hat{C}^{Latt}_{ij}$ & $ \hat T_{ij} $ \\ 
	&	&
	& NAIVE 	& HYP(I)	& HYP(II)/$\overline{\rm Fat7}$
	& 	\\
\hline
$ [ T_{\mu\nu} \times A_\mu ]
	[ T_{\mu\nu} \times A_\mu ] $ 	 & I & 0  & $+0.018$ & $-0.005$ & $-0.007$ & 0 \\
$ [ T_{\mu\nu} \times A_\mu ]		
	[ T_{\mu\nu} \times A_\mu ] $  	 & II & 0 & $-1.061$ & $-0.031$ & $-0.051$ & 0 \\
$ [ T_{\mu\nu} \times A_\mu ]		
	[ T_{\mu\nu} \times A_\nu ] $ 	 & I & 0  & $-0.055$ & $-0.001$ & $-0.007$ & 0 \\
$ [ T_{\mu\nu} \times A_\mu ]		
	[ T_{\mu\nu} \times A_\nu ] $  	 & II & 0 & $-0.191$ & $-0.005$ & $-0.004$ & 0 \\
$ [ T_{\mu\nu} \times A_\mu ]
	[ T_{\mu\nu} \times A_\rho ] $ 	 & I & 0  & $+0.032$ & $+0.001$ & $+0.001$ & 0 \\
$ [ T_{\mu\nu} \times A_\mu ]		
	[ T_{\mu\nu} \times A_\rho ] $   & II & 0 & $-0.096$ & $-0.002$ & $-0.002$ & 0 \\
\hline
\end{tabular}
\end{center}
\end{table}

\clearpage

\begin{table}[ht]
\caption{Renormalization constants for 
${\cal O}_i = ({\cal O}^{Latt}_1)_{II}=
[ V_\mu \times P ][ V_\mu \times P]_{II} + 
[ A_\mu \times P ][ A_\mu \times P]_{II}$.
}
\label{tab:ff-op-1-II-1}
\begin{center}
\begin{tabular}{ l c c c c c c }
${\cal O}^{Latt}_j $ & color trace	& $\hat{\gamma}_{ij}$
	& $\hat{C}^{Latt}_{ij}$ & $\hat{C}^{Latt}_{ij}$ 
	& $\hat{C}^{Latt}_{ij}$ & $ \hat T_{ij} $ \\ 
	&	&
	& NAIVE 	& HYP(I)	& HYP(II)/$\overline{\rm Fat7}$
	& 	\\
\hline
$ [ S \times V_\mu ][ S \times V_\mu ] $ & I 	& 0 & $-5.587$ & $-1.140$ & $-1.058$ & 0 \\
$ [ S \times V_\mu ][ S \times V_\mu ] $ & II 	& 0 & $+1.862$ & $+0.380$ & $+0.353$ & 0 \\
$ [ S \times V_\mu ][ S \times V_\nu ] $ & I 	& 0 & $+1.112$ & $+0.346$ & $+0.312$ & 0 \\
$ [ S \times V_\mu ][ S \times V_\nu ] $ & II 	& 0 & $-0.371$ & $-0.115$ & $-0.104$ & 0 \\
$ [ S \times A_\mu ][ S \times A_\mu ] $ & I 	& 0 & $-1.311$ & $-0.028$ & $-0.028$ & 0 \\
$ [ S \times A_\mu ][ S \times A_\mu ] $ & II 	& 0 & $+0.437$ & $+0.009$ & $+0.009$ & 0 \\
\hline
\end{tabular}
\end{center}
\end{table}
\begin{table}[ht]
\caption{Renormalization constants for 
$({\cal O}^{Latt}_1)_{II}$
(continued from Table~\ref{tab:ff-op-1-II-1}).
}
\label{tab:ff-op-1-II-2}
\begin{center}
\begin{tabular}{ l c c c c c c }
${\cal O}^{Latt}_j $ & color trace	& $\hat{\gamma}_{ij}$
	& $\hat{C}^{Latt}_{ij}$ & $\hat{C}^{Latt}_{ij}$ 
	& $\hat{C}^{Latt}_{ij}$ & $ \hat T_{ij} $ \\ 
	&	&
	& NAIVE 	& HYP(I)	& HYP(II)/$\overline{\rm Fat7}$
	& 	\\
\hline 
$ [ T_{\mu\nu} \times A_\mu ]
	[ T_{\mu\nu} \times A_\mu ] $ & I & 0 		& $-0.772$ & $-0.017$ & $-0.019$ & 0 \\
$ [ T_{\mu\nu} \times A_\mu ]
	[ T_{\mu\nu} \times A_\mu ] $ & II & 0 		& $+0.257$ & $+0.006$ & $+0.006$ & 0 \\
$ [ T_{\mu\nu} \times A_\mu ]
	[ T_{\mu\nu} \times A_\nu ] $ & I & 0 		& $-0.045$ & $-0.001$ & $-0.003$ & 0 \\
$ [ T_{\mu\nu} \times A_\mu ]
	[ T_{\mu\nu} \times A_\nu ] $ & II & 0 		& $+0.015$ & $+0.000$ & $+0.001$ & 0 \\
$ [ T_{\mu\nu} \times A_\rho ]
	[ T_{\mu\nu} \times A_\rho ] $ & I & 0 		& $-1.705$ & $-0.057$ & $-0.112$ & 0 \\
$ [ T_{\mu\nu} \times A_\rho ]
	[ T_{\mu\nu} \times A_\rho ] $ & II & 0 	& $+0.568$ & $+0.019$ & $+0.037$ & 0 \\
\hline
\end{tabular}
\end{center}
\end{table}

\clearpage

\begin{table}[ht]
\caption{Renormalization constants for
${\cal O}_i = ({\cal O}^{Latt}_2)_{I}=
[ V_\mu \times P ][ V_\mu \times P]_{I} - 
[ A_\mu \times P ][ A_\mu \times P]_{I}$.
}
\label{tab:ff-op-2-I-1}
\begin{center}
\begin{tabular}{ l c c c c c c }
${\cal O}^{Latt}_j $ & color trace	& $\hat{\gamma}_{ij}$
	& $\hat{C}^{Latt}_{ij}$ & $\hat{C}^{Latt}_{ij}$ 
	& $\hat{C}^{Latt}_{ij}$ & $ \hat T_{ij} $ \\ 
	&	&
	& NAIVE 	& HYP(I)	& $HYP(II)/\overline{\rm Fat7}$
	& 	\\
\hline								     
$ [ S \times V_\mu ][ S \times V_\mu ] $ & I & 0 	& $+23.854$& $+2.698$ & $+3.103$ & $-1$  \\
$ [ S \times V_\mu ][ S \times V_\mu ] $ & II & 0 	& $+0.842$ & $+0.398$ & $+0.377$ & 0  \\
$ [ S \times V_\mu ][ S \times V_\nu ] $ & I & 0 	& $-0.139$ & $-0.030$ & $-0.024$ & 0  \\
$ [ S \times V_\mu ][ S \times V_\nu ] $ & II & 0 	& $+0.416$ & $+0.089$ & $+0.073$ & 0  \\
$ [ S \times A_\mu ][ S \times A_\mu ] $ & I & 0 	& $-1.443$ & $-0.022$ & $-0.015$ & 0  \\
$ [ S \times A_\mu ][ S \times A_\mu ] $ & II & 0 	& $-0.842$ & $-0.023$ & $-0.033$ & 0  \\
\hline
\end{tabular}
\end{center}
\end{table}
\begin{table}[ht]
\caption{Renormalization constants for
$({\cal O}^{Latt}_2)_{I}$
(continued from Table~\ref{tab:ff-op-2-I-1}).
}
\label{tab:ff-op-2-I-2}
\begin{center}
\begin{tabular}{ l c c c c c c }
${\cal O}^{Latt}_j $ & color trace	& $\hat{\gamma}_{ij}$
	& $\hat{C}^{Latt}_{ij}$ & $\hat{C}^{Latt}_{ij}$ 
	& $\hat{C}^{Latt}_{ij}$ & $ \hat T_{ij} $ \\ 
	&	&
	& NAIVE 	& HYP(I)	& $HYP(II)/\overline{\rm Fat7}$
	& 	\\
\hline								     
$ [ T_{\mu\nu} \times A_\mu ]
	[ T_{\mu\nu} \times A_\mu ] $ & I & 0 		& $+1.820$ & $+0.035$ & $+0.037$ & 0 \\
$ [ T_{\mu\nu} \times A_\mu ]
	[ T_{\mu\nu} \times A_\mu ] $ & II & 0 		& $-0.289$ & $-0.014$ & $-0.032$ & 0 \\
$ [ T_{\mu\nu} \times A_\mu ]
	[ T_{\mu\nu} \times A_\nu ] $ & I & 0 		& $+0.049$ & $+0.001$ & $+0.000$ & 0 \\
$ [ T_{\mu\nu} \times A_\mu ]
	[ T_{\mu\nu} \times A_\nu ] $ & II & 0 		& $-0.146$ & $-0.003$ & $-0.001$ & 0 \\
$ [ T_{\mu\nu} \times A_\mu ]
	[ T_{\mu\nu} \times A_\rho ] $ & I & 0 		& $+0.032$ & $+0.001$ & $+0.001$ & 0 \\
$ [ T_{\mu\nu} \times A_\mu ]
	[ T_{\mu\nu} \times A_\rho ] $ & II & 0 	& $-0.096$ & $-0.002$ & $-0.002$ & 0 \\
\hline
\end{tabular}
\end{center}
\end{table}

\clearpage

\begin{table}[ht]
\caption{Renormalization constants for
${\cal O}_i = ({\cal O}^{Latt}_2)_{II}=
[ V_\mu \times P ][ V_\mu \times P]_{II} - 
[ A_\mu \times P ][ A_\mu \times P]_{II}$.
}
\label{tab:ff-op-2-II-1}
\begin{center}
\begin{tabular}{ l c c c c c c }
${\cal O}^{Latt}_j $ & color trace	& $\hat{\gamma}_{ij}$
	& $\hat{C}^{Latt}_{ij}$ & $\hat{C}^{Latt}_{ij}$ 
	& $\hat{C}^{Latt}_{ij}$ & $ \hat T_{ij} $ \\ 
	&	&
	& NAIVE 	& HYP(I)	& $HYP(II)/\overline{\rm Fat7}$
	& 	\\
\hline
$ [ S \times V_\mu ][ S \times V_\mu ] $ & I & 0 	& $+5.587$ & $+1.140$ & $+1.058$ & 0 \\
$ [ S \times V_\mu ][ S \times V_\mu ] $ & II & 0 	& $-1.862$ & $-0.380$ & $-0.353$ & 0 \\
$ [ S \times V_\mu ][ S \times V_\nu ] $ & I & 0 	& $-1.112$ & $-0.346$ & $-0.312$ & 0 \\
$ [ S \times V_\mu ][ S \times V_\nu ] $ & II & 0 	& $+0.371$ & $+0.115$ & $+0.104$ & 0 \\
$ [ S \times A_\mu ][ S \times A_\mu ] $ & I & 0 	& $-1.311$ & $-0.028$ & $-0.028$ & 0 \\
$ [ S \times A_\mu ][ S \times A_\mu ] $ & II & 0 	& $+0.437$ & $+0.009$ & $+0.009$ & 0 \\
\hline
\end{tabular}
\end{center}
\end{table}
\begin{table}[ht]
\caption{Renormalization constants for
$({\cal O}^{Latt}_2)_{II}$
(continued from Table~\ref{tab:ff-op-2-II-1}).
}
\label{tab:ff-op-2-II-2}
\begin{center}
\begin{tabular}{ l c c c c c c }
${\cal O}^{Latt}_j $ & color trace	& $\hat{\gamma}_{ij}$
	& $\hat{C}^{Latt}_{ij}$ & $\hat{C}^{Latt}_{ij}$ 
	& $\hat{C}^{Latt}_{ij}$ & $ \hat T_{ij} $ \\ 
	&	&
	& NAIVE 	& HYP(I)	& $HYP(II)/\overline{\rm Fat7}$
	& 	\\
\hline
$ [ T_{\mu\nu} \times A_\mu ]
	[ T_{\mu\nu} \times A_\mu ] $ & I & 0 		& $+0.772$ & $+0.017$ & $+0.019$ & 0 \\
$ [ T_{\mu\nu} \times A_\mu ]
	[ T_{\mu\nu} \times A_\mu ] $ & II & 0 		& $-0.257$ & $-0.006$ & $-0.006$ & 0 \\
$ [ T_{\mu\nu} \times A_\mu ]
	[ T_{\mu\nu} \times A_\nu ] $ & I & 0 		& $+0.045$ & $+0.001$ & $+0.003$ & 0 \\
$ [ T_{\mu\nu} \times A_\mu ]
	[ T_{\mu\nu} \times A_\nu ] $ & II & 0 		& $-0.015$ & $+0.000$ & $-0.001$ & 0 \\
$ [ T_{\mu\nu} \times A_\rho ]
	[ T_{\mu\nu} \times A_\rho ] $ & I & 0 		& $+1.705$ & $+0.057$ & $+0.112$ & 0 \\
$ [ T_{\mu\nu} \times A_\rho ]
	[ T_{\mu\nu} \times A_\rho ] $ & II & 0 	& $-0.568$ & $-0.019$ & $-0.037$ & 0 \\
\hline
\end{tabular}
\end{center}
\end{table}

\clearpage

\begin{table}[ht]
\caption{Renormalization constants for
${\cal O}_i=({\cal O}^{Latt}_3)_{I}=
2 [ P \times P ][ P \times P]_{I} - 
2 [ S \times P ][ S \times P]_{I}$.
}
\label{tab:ff-op-3-I-1}
\begin{center}
\begin{tabular}{ l c c c c c c }
${\cal O}^{Latt}_j $ & color trace	& $\hat{\gamma}_{ij}$
	& $\hat{C}^{Latt}_{ij}$ & $\hat{C}^{Latt}_{ij}$ 
	& $\hat{C}^{Latt}_{ij}$ & $ \hat T_{ij} $ \\ 
	&	&
	& NAIVE 	& HYP(I)	& $HYP(II)/\overline{\rm Fat7}$
	& 	\\
\hline								     
$ [ S \times S ][ S \times S ] $ & I & 0 	& $-0.154$ & $-0.005$ & $-0.014$ & 0 \\
$ [ S \times S ][ S \times S ] $ & II & 0 	& $-0.507$ & $-0.011$ & $-0.031$ & 0 \\
$ [ S \times T_{\mu\nu} ]
	[ S \times T_{\mu\nu} ] $ & I & 0 	& $-1.068$ & $-0.004$ & $+0.039$ & 0 \\
$ [ S \times T_{\mu\nu} ]
	[ S \times T_{\mu\nu} ] $ & II & 0 	& $-0.458$ & $-0.002$ & $+0.017$ & 0 \\
$ [ S \times T_{\mu\nu} ]
	[ S \times T_{\mu\rho} ] $ & I & 0 	& $-0.349$ & $-0.018$ & $-0.014$ & 0 \\
$ [ S \times T_{\mu\nu} ]
	[ S \times T_{\mu\rho} ] $ & II & 0 	& $-0.150$ & $-0.008$ & $-0.006$ & 0 \\
\hline
\end{tabular}
\end{center}
\end{table}
\begin{table}[ht]
\caption{Renormalization constants for
$({\cal O}^{Latt}_3)_{I}$
(continued from Table~\ref{tab:ff-op-3-I-2}).
}
\label{tab:ff-op-3-I-2}
\begin{center}
\begin{tabular}{ l c c c c c c }
${\cal O}^{Latt}_j $ & color trace	& $\hat{\gamma}_{ij}$
	& $\hat{C}^{Latt}_{ij}$ & $\hat{C}^{Latt}_{ij}$ 
	& $\hat{C}^{Latt}_{ij}$ & $ \hat T_{ij} $ \\ 
	&	&
	& NAIVE 	& HYP(I)	& $HYP(II)/\overline{\rm Fat7}$
	& 	\\
\hline								     
$ [ A_\mu \times V_\mu ]
	[ A_\mu \times V_\mu ] $ & I & 0 	& $+32.136$& $+3.228$ & $+4.449$ & $-2$ \\
$ [ A_\mu \times V_\mu ]
	[ A_\mu \times V_\mu ] $ & II & 0 	& $+2.578$ & $+0.593$ & $+0.568$ & 0 \\
$ [ A_\mu \times V_\nu ]
	[ A_\mu \times V_\nu ] $ & I & 0 	& $-0.451$ & $+0.163$ & $+0.161$ & 0 \\
$ [ A_\mu \times V_\nu ]
	[ A_\mu \times V_\nu ] $ & II & 0 	& $-0.193$ & $+0.070$ & $+0.069$ & 0 \\
$ [ A_\mu \times V_\mu ]
	[ A_\mu \times V_\nu ] $ & I & 0 	& $+0.304$ & $+0.130$ & $+0.117$ & 0 \\
$ [ A_\mu \times V_\mu ]
	[ A_\mu \times V_\nu ] $ & II & 0 	& $-0.913$ & $-0.391$ & $-0.350$ & 0 \\
\hline
\end{tabular}
\end{center}
\end{table}

\clearpage

\begin{table}[ht]
\caption{Renormalization constants for
${\cal O}_i = ({\cal O}^{Latt}_3)_{II} =
2 [ P \times P ][ P \times P]_{II} - 
2 [ S \times P ][ S \times P]_{II}$.
}
\label{tab:ff-op-3-II-1}
\begin{center}
\begin{tabular}{ l c c c c c c }
${\cal O}^{Latt}_j $ & color trace	& $\hat{\gamma}_{ij}$
	& $\hat{C}^{Latt}_{ij}$ & $\hat{C}^{Latt}_{ij}$ 
	& $\hat{C}^{Latt}_{ij}$ & $ \hat T_{ij} $ \\ 
	&	&
	& NAIVE 	& HYP(I)	& $HYP(II)/\overline{\rm Fat7}$
	& 	\\
\hline
$ [ S \times T_{\mu\nu} ]
	[ S \times T_{\mu\nu} ] $ & I & 0 	& $-0.916$ & $-0.004$ & $+0.033$ & 0 \\
$ [ S \times T_{\mu\nu} ]
	[ S \times T_{\mu\nu} ] $ & II & 0 	& $+0.305$ & $+0.001$ & $-0.011$ & 0 \\
$ [ S \times T_{\mu\nu} ]
	[ S \times T_{\mu\rho} ] $ & I & 0 	& $-0.299$ & $-0.015$ & $-0.012$ & 0 \\
$ [ S \times T_{\mu\nu} ]
	[ S \times T_{\mu\rho} ] $ & II & 0 	& $+0.100$ & $+0.005$ & $+0.004$ & 0 \\
& & &		  	      &			      &  	     	      &			      \\
$ [ S \times P ]
	[ S \times P ] $ & II & $2\times(+16)$ & 
	$2\times(+95.6216)$   & $2\times(+14.6233)$   & $2\times(+19.1180)$ & $2\times(-6 )$  \\
\hline
\end{tabular}
\end{center}
\end{table}

\clearpage

\begin{table}[ht]
\caption{Renormalization constants for
${\cal O}_i = ({\cal O}^{Latt}_4)_{I} =
[ P \times P ][ P \times P]_{I} + 
[ S \times P ][ S \times P]_{I}$.
}
\label{tab:ff-op-S+P-I-1}
\begin{center}
\begin{tabular}{ l c c c c c c }
${\cal O}^{Latt}_j $ & color trace	& $\hat{\gamma}_{ij}$
	& $\hat{C}^{Latt}_{ij}$ & $\hat{C}^{Latt}_{ij}$ 
	& $\hat{C}^{Latt}_{ij}$ & $ \hat T_{ij} $ \\ 
	&	&
	& NAIVE 	& HYP(I)	& $HYP(II)/\overline{\rm Fat7}$
	& 	\\
\hline								     
$ [ S \times S ][ S \times S ] $ & I & 0 	& $-0.077$ & $-0.003$ & $-0.007$ & 0 \\
$ [ S \times S ][ S \times S ] $ & II & 0 	& $-0.254$ & $-0.006$ & $-0.016$ & 0 \\
$ [ S \times T_{\mu\nu} ]
	[ S \times T_{\mu\nu} ] $ & I & 0 	& $-0.977$ & $-0.043$ & $-0.036$ & 0 \\
$ [ S \times T_{\mu\nu} ]
	[ S \times T_{\mu\nu} ] $ & II & 0 	& $-0.419$ & $-0.019$ & $-0.015$ & 0 \\
$ [ S \times T_{\mu\nu} ]
	[ S \times T_{\mu\rho} ] $ & I & 0 	& $-0.099$ & $-0.003$ & $+0.001$ & 0 \\
$ [ S \times T_{\mu\nu} ]
	[ S \times T_{\mu\rho} ] $ & II & 0 	& $-0.042$ & $-0.001$ & $+0.000$ & 0 \\
\hline
\end{tabular}
\end{center}
\end{table}
\begin{table}[ht]
\caption{Renormalization constants for
$({\cal O}^{Latt}_4)_{I}$
(continued from Table~\ref{tab:ff-op-S+P-I-1}).
}
\label{tab:ff-op-S+P-I-2}
\begin{center}
\begin{tabular}{ l c c c c c c }
${\cal O}^{Latt}_j $ & color trace	& $\hat{\gamma}_{ij}$
	& $\hat{C}^{Latt}_{ij}$ & $\hat{C}^{Latt}_{ij}$ 
	& $\hat{C}^{Latt}_{ij}$ & $ \hat T_{ij} $ \\ 
	&	&
	& NAIVE 	& HYP(I)	& $HYP(II)/\overline{\rm Fat7}$
	& 	\\
\hline								     
$ [ A_\mu \times V_\mu ]
	[ A_\mu \times V_\mu ] $ & I & 0 	& $-16.068$& $-1.614$ & $-2.224$ & $+1$ \\
$ [ A_\mu \times V_\mu ]
	[ A_\mu \times V_\mu ] $ & II & 0 	& $-1.289$ & $-0.297$ & $-0.284$ & 0 \\
$ [ A_\mu \times V_\nu ]
	[ A_\mu \times V_\nu ] $ & I & 0 	& $-2.229$ & $-0.326$ & $-0.265$ & 0 \\
$ [ A_\mu \times V_\nu ]						      
	[ A_\mu \times V_\nu ] $ & II & 0 	& $-0.955$ & $-0.140$ & $-0.114$ & 0 \\
$ [ A_\mu \times V_\mu ]
	[ A_\mu \times V_\nu ] $ & I & 0 	& $-0.152$ & $-0.065$ & $-0.058$ & 0 \\
$ [ A_\mu \times V_\mu ]						                  
	[ A_\mu \times V_\nu ] $ & II & 0 	& $+0.457$ & $+0.196$ & $+0.175$ & 0 \\
\hline
\end{tabular}
\end{center}
\end{table}

\clearpage

\begin{table}[ht]
\caption{Renormalization constants for
${\cal O}_i = ({\cal O}^{Latt}_4)_{II} =
[ P \times P ][ P \times P]_{II} + 
[ S \times P ][ S \times P]_{II}$.
}
\label{tab:ff-op-S+P-II-1}
\begin{center}
\begin{tabular}{ l c c c c c c }
${\cal O}^{Latt}_j $ & color trace	& $\hat{\gamma}_{ij}$
	& $\hat{C}^{Latt}_{ij}$ & $\hat{C}^{Latt}_{ij}$ 
	& $\hat{C}^{Latt}_{ij}$ & $ \hat T_{ij} $ \\ 
	&	&
	& NAIVE 	& HYP(I)	& $HYP(II)/\overline{\rm Fat7}$
	& 	\\
\hline
$ [ S \times T_{\mu\nu} ]
	[ S \times T_{\mu\nu} ] $ & I & 0 	& $-0.837$ & $-0.037$ & $-0.031$ & 0 \\
$ [ S \times T_{\mu\nu} ]
	[ S \times T_{\mu\nu} ] $ & II & 0 	& $+0.279$ & $+0.012$ & $+0.010$ & 0 \\
$ [ S \times T_{\mu\nu} ]
	[ S \times T_{\mu\rho} ] $ & I & 0 	& $-0.085$ & $-0.003$ & $+0.001$ & 0 \\
$ [ S \times T_{\mu\nu} ]
	[ S \times T_{\mu\rho} ] $ & II & 0 	& $+0.028$ & $+0.001$ & $+0.000$ & 0 \\
			 &    &	      &		     &		    &		   & 	         \\
$ [ S \times P ]
      [ S \times P ] $ & II & $-16$ & $-95.6216$   & $-14.6233$   & $-19.1180$   & $+6 $ \\
\hline
\end{tabular}
\end{center}
\end{table}

\clearpage

\begin{table}[ht]
\caption{Renormalization constants for
${\cal O}_i = ({\cal O}^{Latt}_5)_{I} =
-\frac12 [ P \times P ][ P \times P]_{I} 
-\frac12 [ S \times P ][ S \times P]_{I}
+\frac12 \sum_{\mu<\nu} [T_{\mu\nu} \times P][T_{\mu\nu} \times P]_I$.
}
\label{tab:ff-op-T-S-P-I-0}
\begin{center}
\begin{tabular}{ l c c c c c c }
${\cal O}^{Latt}_j $ & color trace	& $\hat{\gamma}_{ij}$
	& $\hat{C}^{Latt}_{ij}$ & $\hat{C}^{Latt}_{ij}$ 
	& $\hat{C}^{Latt}_{ij}$ & $ \hat T_{ij} $ \\ 
	&	&
	& NAIVE 	& HYP(I)	& $HYP(II)/\overline{\rm Fat7}$
	& 	\\
\hline								     
$ [ S \times S ]
	[ S \times S ] $ & I & 0 		& $-0.294$ & $-0.003$ & $-0.013$ & 0 \\
$ [ S \times S ]						    
	[ S \times S ] $ & II & 0 		& $-0.016$ & $+0.001$ & $+0.001$ & 0 \\
$ [ S \times T_{\mu\nu} ]					    
	[ S \times T_{\mu\nu} ] $ & I & 0 	& $+2.001$ & $+0.066$ & $+0.027$ & 0 \\
$ [ S \times T_{\mu\nu} ]					    
	[ S \times T_{\mu\nu} ] $ & II & 0 	& $-0.568$ & $-0.038$ & $-0.041$ & 0 \\
$ [ S \times T_{\mu\nu} ]					    
	[ S \times T_{\mu\rho} ] $ & I & 0 	& $+0.037$ & $+0.003$ & $+0.003$ & 0 \\ 
$ [ S \times T_{\mu\nu} ]					    
	[ S \times T_{\mu\rho} ] $ & II & 0 	& $-0.111$ & $-0.010$ & $-0.009$ & 0 \\
\hline
\end{tabular}
\end{center}
\end{table}
\begin{table}[ht]
\caption{Renormalization constants for
$({\cal O}^{Latt}_5)_{I}$
(continued from Table~\ref{tab:ff-op-T-S-P-I-0}).
}
\label{tab:ff-op-T-S-P-I-1}
\begin{center}
\begin{tabular}{ l c c c c c c }
${\cal O}^{Latt}_j $ & color trace	& $\hat{\gamma}_{ij}$
	& $\hat{C}^{Latt}_{ij}$ & $\hat{C}^{Latt}_{ij}$ 
	& $\hat{C}^{Latt}_{ij}$ & $ \hat T_{ij} $ \\ 
	&	&
	& NAIVE 	& HYP(I)	& $HYP(II)/\overline{\rm Fat7}$
	& 	\\
\hline								     
$ [ T_{\mu\nu} \times S ]
	[ T_{\mu\nu} \times S ] $ & I & 0 		& $+0.358$ & $+0.004$ & $+0.017$ & 0 \\
$ [ T_{\mu\nu} \times S ]
	[ T_{\mu\nu} \times S ] $ & II & 0 		& $-0.174$ & $-0.003$ & $-0.010$ & 0 \\
$ [ T_{\mu\nu} \times T_{\mu\nu} ]
	[ T_{\mu\nu} \times T_{\mu\nu} ] $ & I & 0 	& $-1.761$ & $-0.055$ & $-0.023$ & 0 \\
$ [ T_{\mu\nu} \times T_{\mu\nu} ]
	[ T_{\mu\nu} \times T_{\mu\nu} ] $ & II & 0 	& $-0.151$ & $+0.007$ & $+0.030$ & 0 \\
$ [ T_{\mu\nu} \times T_{\mu\rho} ]
	[ T_{\mu\nu} \times T_{\mu\rho} ] $ & I & 0 	& $-1.602$ & $-0.039$ & $+0.004$ & 0 \\
$ [ T_{\mu\nu} \times T_{\mu\rho} ]
	[ T_{\mu\nu} \times T_{\mu\rho} ] $ & II & 0 	& $-0.627$ & $-0.041$ & $-0.051$ & 0 \\
\hline
\end{tabular}
\end{center}
\end{table}
\begin{table}[ht]
\caption{Renormalization constants for
$({\cal O}^{Latt}_5)_{I}$
(continued from Table~\ref{tab:ff-op-T-S-P-I-1}).
}
\label{tab:ff-op-T-S-P-I-2}
\begin{center}
\begin{tabular}{ l c c c c c c }
${\cal O}^{Latt}_j $ & color trace	& $\hat{\gamma}_{ij}$
	& $\hat{C}^{Latt}_{ij}$ & $\hat{C}^{Latt}_{ij}$ 
	& $\hat{C}^{Latt}_{ij}$ & $ \hat T_{ij} $ \\ 
	&	&
	& NAIVE 	& HYP(I)	& $HYP(II)/\overline{\rm Fat7}$
	& 	\\
\hline								     
$ [ A_\mu \times V_\mu ]
	[ A_\mu \times V_\mu ] $ & I & 0 	& $+12.330$& $+1.436$ & $+1.624$ & $-1/2$ \\
$ [ A_\mu \times V_\mu ]
	[ A_\mu \times V_\mu ] $ & II & 0 	& $-0.788$ & $-0.061$ & $-0.029$ & 0 \\
$ [ A_\mu \times V_\nu ]
	[ A_\mu \times V_\nu ] $ & I & 0 	& $+12.483$& $+1.490$ & $+1.685$ & $-1/2$ \\
$ [ A_\mu \times V_\nu ]
	[ A_\mu \times V_\nu ] $ & II & 0 	& $-1.241$ & $-0.225$ & $-0.212$ & 0 \\
$ [ A_\mu \times V_\mu ]
	[ A_\mu \times V_\nu ] $ & I & 0 	& $+0.076$ & $+0.033$ & $+0.029$ & 0 \\
$ [ A_\mu \times V_\mu ]
	[ A_\mu \times V_\nu ] $ & II & 0 	& $-0.228$ & $-0.098$ & $-0.087$ & 0 \\
\hline
\end{tabular}
\end{center}
\end{table}

\clearpage

\begin{table}[ht]
\caption{Renormalization constants for
${\cal O}_i = ({\cal O}^{Latt}_5)_{II} =
-\frac12 [ P \times P ][ P \times P]_{II}
-\frac12 [ S \times P ][ S \times P]_{II}
+\frac12 \sum_{\mu<\nu} [T_{\mu\nu}\times P][T_{\mu\nu}\times P]_{II}$.
}
\label{tab:ff-op-T-S-P-II-0}
\begin{center}
\begin{tabular}{ l c c c c c c }
${\cal O}^{Latt}_j $ & color trace	& $\hat{\gamma}_{ij}$
	& $\hat{C}^{Latt}_{ij}$ & $\hat{C}^{Latt}_{ij}$ 
	& $\hat{C}^{Latt}_{ij}$ & $ \hat T_{ij} $ \\ 
	&	&
	& NAIVE 	& HYP(I)	& $HYP(II)/\overline{\rm Fat7}$
	& 	\\
\hline
$ [ S \times S ]
	[ S \times S ] $ & I & 0 		& $-0.286$ & $-0.004$ & $-0.015$ & 0 \\
$ [ S \times S ]						    
	[ S \times S ] $ & II & 0 		& $+0.095$ & $+0.001$ & $+0.005$ & 0 \\
$ [ S \times T_{\mu\nu} ]					    
	[ S \times T_{\mu\nu} ] $ & I & 0 	& $+0.419$ & $+0.019$ & $+0.015$ & 0 \\
$ [ S \times T_{\mu\nu} ]					    
	[ S \times T_{\mu\nu} ] $ & II & 0 	& $-0.140$ & $-0.006$ & $-0.005$ & 0 \\
$ [ S \times T_{\mu\nu} ]					    
	[ S \times T_{\mu\rho} ] $ & I & 0 	& $+0.042$ & $+0.001$ & $+0.000$ & 0 \\
$ [ S \times T_{\mu\nu} ]					    
	[ S \times T_{\mu\rho} ] $ & II & 0 	& $-0.014$ & $+0.000$ & $+0.000$ & 0 \\
\hline
\end{tabular}
\end{center}
\end{table}
\begin{table}[ht]
\caption{Renormalization constants for
$({\cal O}^{Latt}_5)_{II}$
(continued from Table~\ref{tab:ff-op-T-S-P-II-0}).
}
\label{tab:ff-op-T-S-P-II-1}
\begin{center}
\begin{tabular}{ l c c c c c c }
${\cal O}^{Latt}_j $ & color trace	& $\hat{\gamma}_{ij}$
	& $\hat{C}^{Latt}_{ij}$ & $\hat{C}^{Latt}_{ij}$ 
	& $\hat{C}^{Latt}_{ij}$ & $ \hat T_{ij} $ \\ 
	&	&
	& NAIVE 	& HYP(I)	& $HYP(II)/\overline{\rm Fat7}$
	& 	\\
\hline
$ [ T_{\mu\nu} \times S ]
	[ T_{\mu\nu} \times S ] $ & I & 0 		& $+0.095$ & $+0.001$ & $+0.005$ & 0 \\
$ [ T_{\mu\nu} \times S ]
	[ T_{\mu\nu} \times S ] $ & II & 0 		& $-0.032$ & $+0.000$ & $-0.002$ & 0 \\
$ [ T_{\mu\nu} \times T_{\mu\nu} ]
	[ T_{\mu\nu} \times T_{\mu\nu} ] $ & I & 0 	& $-1.335$ & $-0.059$ & $-0.045$ & 0 \\
$ [ T_{\mu\nu} \times T_{\mu\nu} ]
	[ T_{\mu\nu} \times T_{\mu\nu} ] $ & II & 0 	& $+0.445$ & $+0.020$ & $+0.015$ & 0 \\
$ [ T_{\mu\nu} \times T_{\mu\rho} ]
	[ T_{\mu\nu} \times T_{\mu\rho} ] $ & I & 0 	& $-1.712$ & $-0.102$ & $-0.117$ & 0 \\
$ [ T_{\mu\nu} \times T_{\mu\rho} ]					
	[ T_{\mu\nu} \times T_{\mu\rho} ] $ & II & 0 	& $+0.571$ & $+0.034$ & $+0.039$ & 0 \\
\hline
\end{tabular}
\end{center}
\end{table}
\begin{table}[ht]
\caption{Renormalization constants for
$({\cal O}^{Latt}_5)_{II}$
(continued from Table~\ref{tab:ff-op-T-S-P-II-1}).
}
\label{tab:ff-op-T-S-P-II-2}
\begin{center}
\begin{tabular}{ l c c c c c c }
${\cal O}^{Latt}_j $ & color trace	& $\hat{\gamma}_{ij}$
	& $\hat{C}^{Latt}_{ij}$ & $\hat{C}^{Latt}_{ij}$ 
	& $\hat{C}^{Latt}_{ij}$ & $ \hat T_{ij} $ \\ 
	&	&
	& NAIVE 	& HYP(I)	& $HYP(II)/\overline{\rm Fat7}$
	& 	\\
\hline
$ [ A_\mu \times V_\mu ]
	[ A_\mu \times V_\mu ] $ & I & 0 	& $+1.506$ & $+0.308$ & $+0.307$ & 0 \\
$ [ A_\mu \times V_\mu ]
	[ A_\mu \times V_\mu ] $ & II & 0 	& $-0.502$ & $-0.103$ & $-0.102$ & 0 \\
$ [ A_\mu \times V_\nu ]
	[ A_\mu \times V_\nu ] $ & I & 0 	& $+0.955$ & $+0.140$ & $+0.114$ & 0 \\
$ [ A_\mu \times V_\nu ]					    
	[ A_\mu \times V_\nu ] $ & II & 0 	& $-0.318$ & $-0.047$ & $-0.038$ & 0 \\
$ [ A_\mu \times V_\mu ]					    
	[ A_\mu \times V_\nu ] $ & I & 0 	& $-0.382$ & $-0.109$ & $-0.096$ & 0 \\
$ [ A_\mu \times V_\mu ]					    
	[ A_\mu \times V_\nu ] $ & II & 0 	& $+0.127$ & $+0.036$ & $+0.032$ & 0 \\
\hline
\end{tabular}
\end{center}
\end{table}

\end{document}